\documentclass[preprint,12pt]{elsarticle}

\usepackage{amssymb}

\usepackage{graphicx}
\usepackage{amsmath}
\usepackage[version=4]{mhchem}
\usepackage{siunitx}
\usepackage{longtable,tabularx}

\usepackage{xcolor}
\usepackage{booktabs}

\usepackage{caption}
\usepackage{subcaption}

\usepackage{bm}

\usepackage{mathrsfs}

\newcommand{\grad}{\mbox{$\nabla$}}

\usepackage[hidelinks]{hyperref}

\def\chapterautorefname~#1\null{Chapter~#1\null}
\def\sectionautorefname~#1\null{Section~#1\null}
\def\subsectionautorefname~#1\null{Section~#1\null}
\def\subsubsectionautorefname~#1\null{Section~#1\null}
\def\equationautorefname~#1\null{Eq.~#1\null}
\def\figureautorefname~#1\null{Fig.~#1\null}	
\def\tableautorefname~#1\null{Table~#1\null}

\journal{Computers \& Fluids}

\begin{document}

\newcommand{\mpp}{Mutation$^{++}$}

\begin{frontmatter}

\title{Assessment of Immersed Boundary Methods for Hypersonic Flows with Gas-Surface Interactions}

\author[label1]{Ata Onur Başkaya} % 

\author[label2,label3]{Michele Capriati} 

\author[label2,label4]{Alessandro Turchi} 

\author[label2]{\\Thierry Magin} % 

\author[label1]{Stefan Hickel} % 

\affiliation[label1]{organization={Aerodynamics Group, Faculty of Aerospace Engineering, TU Delft},%Department and Organization
            country={The Netherlands}}

\affiliation[label2]{organization={Aeronautics and Aerospace Department, von Karman Institute for Fluid Dynamics},%Department and Organization
            country={Belgium}}

\affiliation[label3]{organization={Inria, Centre de Mathématiques Appliquées, Ecole Polytechnique, IPP},%Department and Organization
            country={France}}

\affiliation[label4]{organization={Science and Research Directorate, Italian Space Agency, },%Department and Organization
            country={Italy}}
            
\begin{abstract}
Immersed boundary (IB) methods with adaptive mesh refinement (AMR) techniques are assessed for atmospheric entry applications, including effects of chemical nonequilibrium (CNE) and gas\nobreakdash-surface interactions (GSI).
The performance of a conservative cut-cell and two non-conservative ghost-cell IB methods is assessed in comparison with analytical solutions, data from literature, and results obtained with a reference solver that operates on body\nobreakdash-fitted grids. 
All solvers use the same external thermochemistry library so that all observed differences can be attributed to the underlying numerical methods.
Results from eight benchmark cases are reported. Four cases are selected to verify the implementation of chemistry, transport properties, catalytic boundary conditions, and shock capturing.
Four validation cases consider blunt geometries with adiabatic/isothermal and inert/catalytic/ablative boundary conditions.
Overall, the results obtained with the IB solvers are in very good agreement with the reference data.
Discrepancies arise with ghost\nobreakdash-cell methods for cases with large temperature or concentration gradients at the wall and are attributed to mass conservation errors. Only a strictly conservative cut-cell IB method is on par with body\nobreakdash-fitted grid methods.

\end{abstract}

\begin{graphicalabstract}

\vspace{1.5cm}
\includegraphics[clip, trim=0.45cm 0.255cm 13cm 3.25cm, 
		width=0.2965\columnwidth]{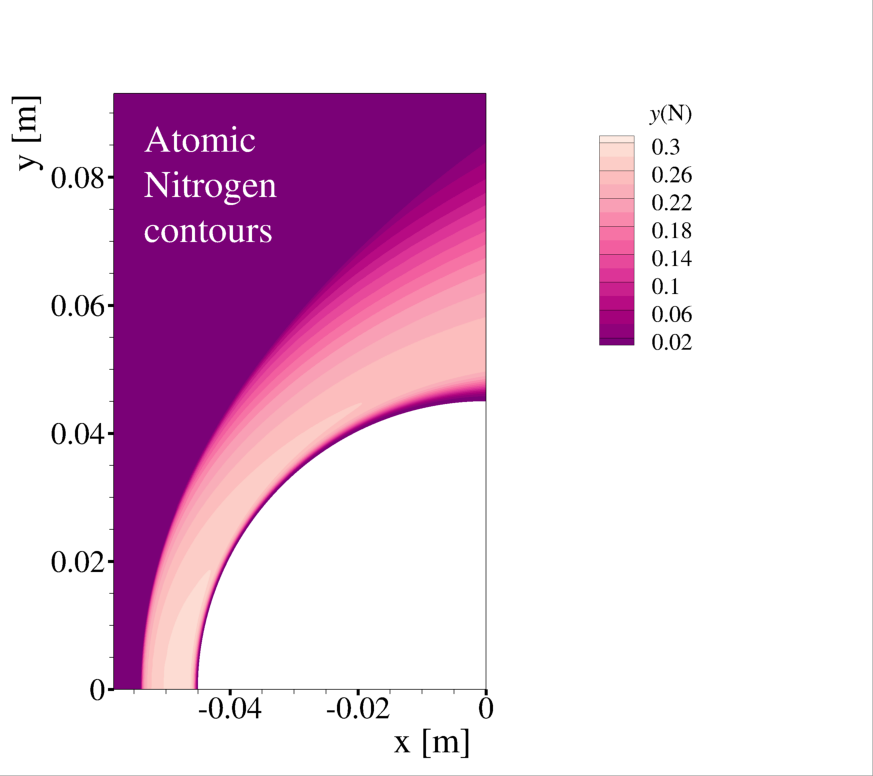}
  \hspace{5mm}
\includegraphics[clip, trim=0cm 0.1cm 0cm 0cm, 
		width=0.55\columnwidth]{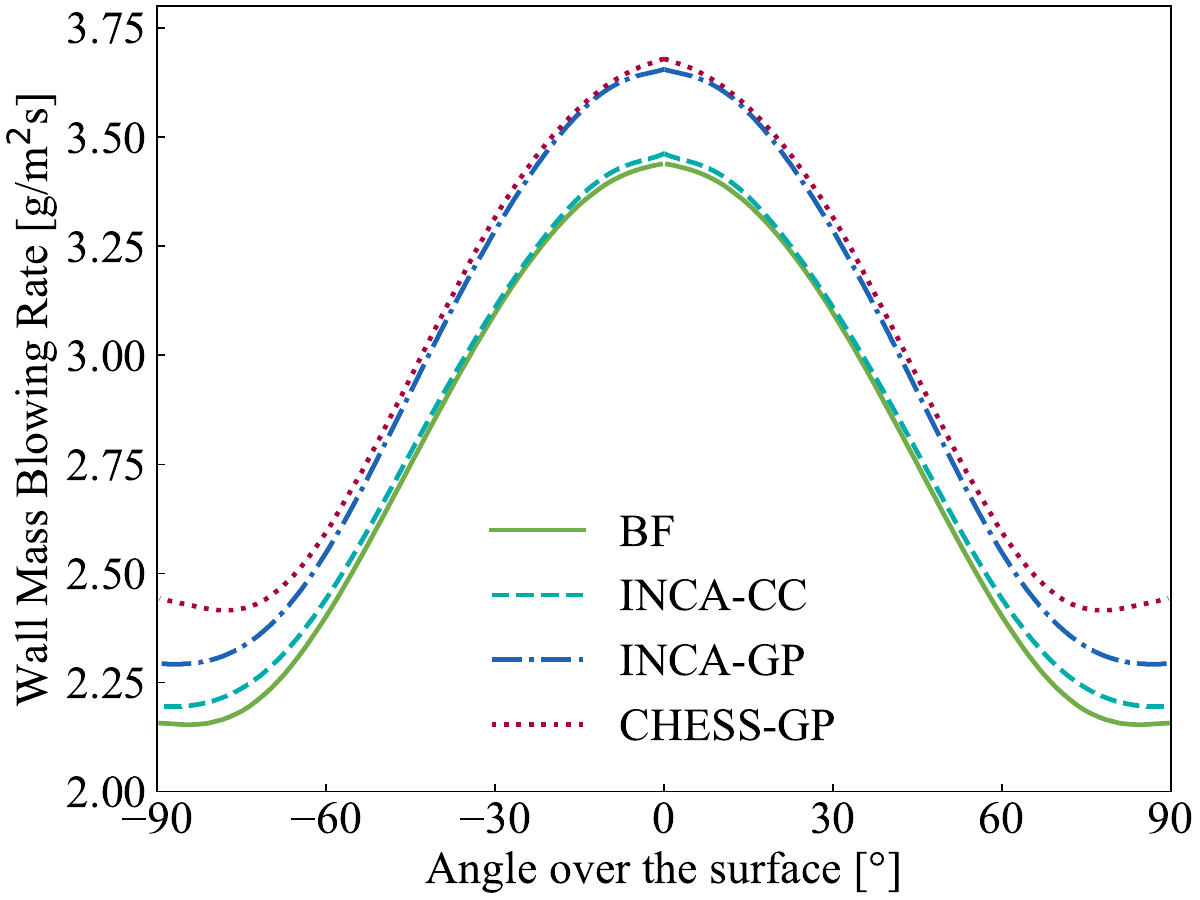}
\end{graphicalabstract}

\begin{highlights}
\item Immersed boundary (IB) methods are assessed for applications with strong thermal gradients and gas-surface interactions.
\item A set of well-defined test cases is established for the verification and validation of IB methods for atmospheric entry.
\item Ghost-cell based IB methods suffer from conservation errors at cold isothermal walls and reacting ablative walls.
\item Strictly conservative cut-cell IB method performs on par with body\nobreakdash-fitted grid methods.
\end{highlights}


\end{frontmatter}


\section{Introduction}

Hypersonic flows experienced during atmospheric entry of capsules or space debris are characterized by strong shock waves and thermochemical nonequilibrium effects through the excitation of the internal energy modes of species and rapid chemical reactions in the shock layer. The hot gas interacts with the surface thermal protection system (TPS) material installed to protect the spacecraft from this hostile environment. Depending on the characteristics of the TPS material, these gas-surface interactions (GSI) involve catalysis as well as ablation.
While the former accelerates the exothermic recombination reactions leading to increased heat transfer towards the surface, the latter alleviates the heat load by means of physicochemical decomposition and mass loss. 
These ablative GSI change the shape of the object by surface recession.
Understanding these interactions is crucial for predicting the surface stresses and heat fluxes, as well as the
uncontrolled trajectory of space debris.
Ground testing is indispensable for validation purposes; however, no facility can simultaneously replicate all aspects of atmospheric entry flows~\cite{gu2020capabilities}.
Hence, computational fluid dynamics (CFD) simulations are essential for the efficient aerothermodynamic analysis and design of future spacecraft.

Most CFD solvers used for high\nobreakdash-speed and high\nobreakdash-enthalpy applications employ body\nobreakdash-fitted structured grids~\cite{wright2009data,knight2012assessment,knight2017assessment}. In these solvers, alignment of the grid
with the shock and the surface needs to be ensured for an
accurate prediction of the flow field. Generating these types of grids usually involves strenuous effort from the user especially for detailed features and incremental geometry updates~\cite{candler2009current}.
Unstructured grids have also been explored; however, issues affecting the heat flux predictions at the surface were reported~\cite{candler2015development,scalabrin2005development}.

A promising alternative is the use of adaptive mesh refinement (AMR) techniques based on piecewise Cartesian grids with immersed boundary (IB) methods.
There has been a recently growing interest in IB\nobreakdash-AMR solvers for atmospheric entry applications~\cite{arslanbekov2011analysis, sekhar2013predictions, atkins2020towards, mcquaid2021immersed, brahmachary2021role},
mainly for their potential in considering complex and deforming geometries, and better robustness and higher computational efficiency compared to body\nobreakdash-fitted mesh-deformation methods. 
These methods also allow for a relatively straightforward implementation of high-order schemes.
However, special care must be taken to have sufficient grid resolution near the boundaries, as it is more difficult for immersed boundary methods to efficiently resolve thin boundary layers over curved surfaces.
To address this shortcoming, a blend of Cartesian grids in the fluid and body\nobreakdash-fitted grids near the surface can be employed~\cite{atkins2020towards,mcquaid2021immersed}. 
This approach has been successful in reducing the required number of cells and providing better resolution of the thermal boundary layer. In general, a blended grid approach is well suited for shapes with smooth curvatures. 
However, it is susceptible to the same drawbacks inherent to body\nobreakdash-fitted grids, for instance, their difficult adaptation to complex deforming geometries.

Arslanbekov et al.~\cite{arslanbekov2011analysis}, Sekhar and Ruffin~\cite{sekhar2013predictions}, and more recently Brahmachary et al.~\cite{brahmachary2021role} demonstrated the benefits of using IB\nobreakdash-AMR solvers for a number of relevant cases. These studies have generally indicated good predictions for wall pressure and skin friction distributions, while emphasizing the difficulty in accurately predicting wall heat fluxes.
As with more recent contributions~\cite{atkins2020towards, mcquaid2021immersed}, these studies were mostly performed with ghost\nobreakdash-cell methods from the family of discrete forcing IB approaches~\cite{mittal2005immersed}.
Ghost-cell methods impose boundary conditions by extrapolating the fluid solution into the solid, i.e. into ghost cells that are neighbouring fluid cells.
Relying solely on ghost\nobreakdash-point extrapolation does not ensure strict conservation of mass, momentum, and energy.
A strictly conservative approach is the cut\nobreakdash-cell finite-volume method, which splits fluid and solid domains into consistently deformed finite volumes.
The implementation of a cut\nobreakdash-cell method for three dimensions and high-order schemes is not as straightforward as the ghost\nobreakdash-cell approach, and it also introduces additional challenges such as cut-cells with very small fluid volumes. 
However, the main advantage of the cut-cell method lies in satisfying the conservation of mass, momentum, and energy near the wall~\cite{mittal2005immersed}.

In this paper, ghost-cell and cut-cell IB methods are scrutinized through a curated list of benchmark case studies relevant for atmospheric entry applications. Main contributions of this paper are twofold:
\begin{itemize}
    \item to assess the accuracy of IB methods for applications with strong thermal gradients and gas-surface interactions.
    \item to establish a set of well-defined test cases for the verification and validation of IB methods for atmospheric entry.
\end{itemize}
Results obtained with the IB-AMR solvers INCA~\cite{hickel2014subgrid,muller2016large} and CHESS~\cite{baskaya2022verification,ninni2022phd} are compared to reference results obtained with the body\nobreakdash-fitted finite\nobreakdash-volume solver US3D~\cite{candler2015development} in addition to data from literature.
A consistent comparison to assess the accuracy of the numerical methods is achieved by coupling each of the flow solvers with the same thermochemistry library, \mpp~\cite{scoggins2020mutation}.
The paper is structured as follows:
Governing equations and modelling approaches are presented in~\autoref{sec:governing}. Solver methodologies are introduced in~\autoref{sec:num_meth}. Results of the benchmark case studies are presented and discussed in~\autoref{sec:results}, while the influence of the different IB methologies is further investigated in~\autoref{sec:GP_vs_CC}. Concluding remarks are made in~\autoref{sec:conc}.

\section{Governing Equations and Models} \label{sec:governing}

\subsection{Governing Equations}

The compressible Navier-Stokes equations are solved in their conservative form for a reacting multicomponent fluid,
\begin{align}
\frac{\partial \rho_i}{\partial t} + \bm{\grad} \cdot \left( \rho_i \mathbf{u} \right) + \bm{\grad} \cdot \mathbf{J}_i &= \dot{\omega}_i \:, \label{eq:gov_1} \\
\frac{\partial \rho \mathbf{u}}{\partial t} 
+ \bm{\grad} \cdot \left( \rho \mathbf{u} \otimes \mathbf{u} \right) + \bm{\grad} p - \bm{\grad} \cdot \bm{\tau}
&= 0 \:, \label{eq:gov_2} \\
\frac{\partial \rho E}{\partial t} + \bm{\grad} \cdot \left[ \left( \rho E + p \right) \mathbf{u} \right] + \bm{\grad} \cdot \mathbf{q} - \bm{\grad} \cdot \left( \bm{\tau} \cdot \mathbf{u} \right) &= 0 \label{eq:gov_3} \:,
\end{align}
where $ \rho_i $ is the species partial density for the $ i^\text{th} $ species, $ \mathbf{u} $ is the mixture average velocity, $ \dot{\omega}_i $ is the source term associated with the production or consumption of species due to chemical reactions, $ \rho $ is the mixture density, $ p $ is the mixture pressure, and $ E = e + u^2 / 2 $ is the specific total energy, which is the sum of the thermodynamic internal energy $ e $ and the kinetic energy.
External forces due to gravitational or electromagnetic effects, and radiative energy exchanges are not considered for the cases in this study. 
Both solvers considered in this work can perform under thermal nonequilibrium with multi-temperature methods, such as that of Park~\cite{park1989nonequilibrium}. However, results presented in this paper are obtained with a thermal equilibrium assumption.

The ideal gas assumption leads to the equation of state $ p = \rho R T $, where $ R = \mathcal{R}/\overline{M} $ is the mixture gas constant obtained from the universal gas constant $ \mathcal{R} $ and the mixture average molar mass $ \overline{M} $.
These mixture properties are modelled according to Dalton's law through their constituent species as $ p = \sum_i p_i$, $ \rho = \sum_i \rho_i $, $ R = \sum_i y_i R_i $, with the mass fractions $ y_i = \rho_i / \rho $.

Two models are considered for the species diffusion flux $ \mathbf{J}_i $: Fick's law with a correction to ensure conservation of mass as
\begin{equation} \label{eq:diff_Jmassgrad}
\mathbf{J}_i = - \rho D_{im}  \bm{\grad}y_i + y_i \sum_j \rho D_{jm}  \bm{\grad}y_j \:,
\end{equation}
with the mixture-averaged diffusion coefficients $ D_{im} = \frac{1-x_i}{\sum_{j \neq i} \frac{x_j}{\mathscr{D}_{ij} }} \:, $ obtained by Wilke's average of the binary diffusion coefficients $ \mathscr{D}_{ij} $.
The second diffusion model uses the solution of the Stefan\nobreakdash-Maxwell equations,
\begin{equation} \label{eq:s-m-original}
\bm{\grad} x_i 
	= \frac{\overline{M}}{\rho} \sum_{j \ne i} \left( 
	  \frac{x_i \mathbf{J}_j }{M_j \mathscr{D}_{ij}} - \frac{x_j \mathbf{J}_i }{M_i \mathscr{D}_{ij}} \right) \: ,
\end{equation}
where $ x_i $ are the mole fractions, and $ M_i $ are the species molar masses.
This formulation is computationally costlier, but theoretically more accurate~\cite{sutton1998multi}.

Viscosity and thermal conductivity are obtained through a linear system solution using an LDL$^\text{T}$ decomposition as opposed to the common approach of using simplified mixture rules~\cite{wilke1950viscosity,mason1958approximate,sarma2000physico}.
The viscous stress tensor $ \bm{\tau} $ is defined assuming Stokes' hypothesis as
\begin{equation}
\bm{\tau} = \mu \left[ \bm{\grad} \mathbf{u} + \left(\bm{\grad} \mathbf{u}\right)^\dagger  - \frac{2}{3} \bm{\grad} \cdot \mathbf{u} \bm{I} \right] \: ,
\end{equation}
where $ \mu $ is the dynamic (shear) viscosity of the mixture.

The total heat flux vector $ \mathbf{q} $ includes the contributions from conduction and mass diffusion,
\begin{equation}
\mathbf{q} = -\lambda \bm{\grad} T + \sum_{i} \mathbf{J}_i h_i(T) \: ,
\end{equation}
where $T$ is the temperature. The first term stems from Fourier's law with the thermal conductivity $ \lambda $ of the mixture, and the second term accounts for the transport of enthalpy by species diffusion, with $ h_i $ as the species enthalpy.

\subsection{Physicochemical Modelling}
\label{sec:physico}

The models used in state-of-the-art CFD solvers capable of simulating the aforementioned phenomena vary considerably.
Broadly, choices need to be made on the thermodynamic database, the treatment of TCNE effects, the transport properties modelling, and the approach for handling GSI.
For details we refer to several published studies that evaluate the impact of these selections in modelling thermal nonequilibrium~\cite{park2010limits,magin2012coarse}, species diffusion~\cite{sutton1998multi,alkandry2013comparison}, viscosity and thermal conductivity~\cite{alkandry2013comparison,miro2019high}, rate of catalysis~\cite{hollis2013assessment}, and ablation~\cite{candler2012nonequilibrium}.
As important quantities of interest, such as surface heat fluxes, are highly sensitive to the modelling approaches, large discrepancies between the results obtained with hypersonic CFD codes are common~\cite{knight2012assessment,knight2017assessment}.
Hence, this variety of approaches often obscures a clear assessment of the underlying numerical methods, when comparing different solvers.

Based on these considerations, each of the solvers used in this study is coupled with the multicomponent thermodynamic and transport properties for ionized gases in C++ (\mpp) open-source library. \mpp\ provides all required physiochemical models for thermodynamics, transport properties, chemical kinetics, and GSI.
A detailed description of \mpp~is presented by Scoggins et al.~\cite{scoggins2020mutation}.

The caloric properties of the species are approximated with standard NASA-9 polynomial fits~\cite{mcbride2002nasa}.
Species mass diffusivities, viscosities, and thermal conductivities are provided by \mpp~according to multicomponent Chapman-Enskog formulations~\cite{chapman1990mathematical}.
The chemical reaction mechanisms, that is,  species mass rates and their analytical Jacobians with respect to species densities, are also provided by \mpp. 

Catalytic and ablative surface boundary conditions are imposed by solving a mass balance~\cite{bellas2017development,scoggins2020mutation},
\begin{equation} \label{eq:mass_balance}
    (\rho_iv_{blow})_{wall} + (J_i)_{wall} = \dot{\omega}_{i,wall} \:,
\end{equation}
with $v_{blow}$ as the surface-normal blowing velocity, which is nonzero only for an ablative boundary. 
Terms from left to right refer to convective flux due to blowing, diffusive flux, and species source term due to surface reactions.
A probability based approach is employed for computing this chemical source term for the surface, written as
\begin{equation}\label{eq:react_gamma}
    \dot{\omega}_{i,wall} = \gamma m_i \mathcal{F}_{i,impin} \:,
\end{equation}
where $ \gamma = \mathcal{F}_{i,react} / \mathcal{F}_{i,impin} $ is the ratio of reacting to impinging species fluxes and it describes the efficiency of the process, and $ m_i $ is the mass of the $i^\text{th}$ species~\cite{bellas2017development}.
Assuming the species at the wall have a Maxwellian distribution function, the impinging species flux is
\begin{equation}
    \mathcal{F}_{i,impin} = n_i \sqrt{ \frac{k_B T_{w}}{2 \pi m_i} } \:,
\end{equation}
where $ k_B $ is the Boltzmann constant, $ T_w $ is the wall temperature, and $ n_i $ is the number density of the $i^\textbf{th}$ species~\cite{barbante2001accurate}.
From the mass blowing rate $ \dot{m} = \sum_{i}\dot{\omega}_{i,wall} $, the blowing speed is calculated by
\begin{equation} \label{eq:vblow}
v_{blow} = \frac{\dot{m}}{\sum_{i}\rho_i} \:.
\end{equation}
Values obtained for species densities and mass blowing speeds are then imposed as boundary conditions for the Navier-Stokes equations.

\section{Numerical Methods} 
\label{sec:num_meth}

\begin{figure} [tb]
	\begin{subfigure}[t]{0.325\textwidth}\centering
		\includegraphics[clip, trim=4.5cm 13.5cm 13cm 11.5cm, width=.99\textwidth]{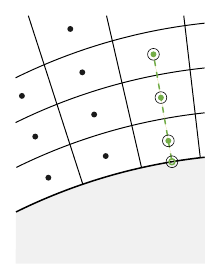}
		\caption{}
        \label{fig:method_sketch_1}
	\end{subfigure}
	\begin{subfigure}[t]{.325\textwidth}\centering
		\includegraphics[clip, trim=8.5cm 13.5cm 9cm 11.5cm, width=.99\textwidth]{IB_sketch_new.pdf}
		\caption{}
        \label{fig:method_sketch_2}
	\end{subfigure}
	\begin{subfigure}[t]{.325\textwidth}\centering
		\includegraphics[clip, trim=12.5cm 13.5cm 5cm 11.5cm, width=.99\textwidth]{IB_sketch_new.pdf}
		\caption{}
        \label{fig:method_sketch_3}
	\end{subfigure}
	\caption{Schematics of (a) body-fitted, (b) ghost-cell IB, and (c) cut-cell IB approaches. Ghost cells are striped (violet or blue) and cut-cells are tinted (turquoise).}
	\label{fig:method_sketch}
\end{figure}

We consider three different methodologies for imposing surface boundary conditions in the framework of finite\nobreakdash-volume methods.
Schematics of the body-fitted, ghost-cell IB, and cut-cell IB approaches are shown in~\autoref{fig:method_sketch}. The arc near the middle of each sketch indicates the surface that demarcates the fluid above from the solid below it. The other black lines are grid lines and the filled dots indicate cell centers. The color code matches the one used for presenting the results in~\autoref{sec:results}.

The classical body-fitted grid method,~\autoref{fig:method_sketch_1}, simply makes use of the grid's alignment with the geometry. An example stencil is drawn in the sketch and the boundary intercept is indicated by a colored hollow circle. Hollow circles indicate the stencil of the discretization scheme.
The fluid-cell solutions and boundary conditions are used to reconstruct quantities at cell interfaces according to the chosen numerical scheme.

The two other approaches use IB methods on Cartesian grids.
The ghost-cell IB approach~\cite{mittal2008versatile} seen in ~\autoref{fig:method_sketch_2} imposes boundary conditions by extending the fluid solution to ghost-cells. The virtual flow solution of ghost-cells is set by extrapolating the nearby fluid solution according to boundary conditions at the nearest fluid-solid interface. Since the grid does not conform with the geometry, the solution at an image point in the fluid needs to be found through interpolation using the surrounding fluid-cell solutions. An example stencil is colored in the sketch, with the fluid points used in the interpolation connected by dotted lines. Ghost-cells are marked with a striped pattern in the sketch.
While being relatively straightforward to implement, this approach does not ensure strict conservation of mass, momentum, and energy at the interface between the fluid and the solid. Fluxes are reconstructed from the fluid-cell and the ghost-cell solutions on the Cartesian grid without considering the location and shape of the fluid-solid interface. Errors in implicitly satisfying the conservative flux boundary condition therefore result from the nonlinearity of the flux function, from the image point interpolation, and from the interface curvature.

The cut-cell IB approach~\cite{ye1999accurate,udaykumar2001sharp,meyer2010conservative}, see~\autoref{fig:method_sketch_3}, ensures strict conservation by considering the flux balance for the part of the cell that belongs to the fluid domain. These consistently deformed finite volumes and their cell faces are colored in the sketch.
Fluxes over the cell faces of the cut cells are scaled according to the wetted areas. The exchange of mass (e.g. with surface reactions), momentum, and energy through the fluid-solid interface is calculated from the prescribed boundary conditions and the local fluid solution.
The latter is acquired by interpolation from the surrounding cell values and the boundary conditions.
An example stencil is colored in the sketch for the cut-cell interpolation.
The other stencil in the sketch is identical to the ghost-cell IB approach. This addition to the cut-cell method refers to the specific implementation within the INCA solver and will be discussed in~\autoref{sec:INCA}. 
Cut-cells with a very small fluid volume fraction require a special treatment to ensure stable time integration. They are typically mixed or merged with nearby cells~\cite{meyer2010conservative}.

\subsection{Body-fitted Solver}
\label{US3D}
The body-fitted solver considered in this study is US3D, which is a high-fidelity flow solver specifically designed for aerodynamic applications in the hypersonic regime by the University of Minnesota and NASA~\cite{candler2015development}.
It solves the compressible chemically reacting Navier\nobreakdash-Stokes equations in a finite\nobreakdash-volume framework on unstructured body\nobreakdash-fitted grids. 
Among the several numerical schemes available in the solver, all simulations carried out within this work use the modified Steger-Warming scheme~\cite{steger1981flux}, which is suitable for steady computations.
A MUSCL approach~\cite{van1979towards} is employed to obtain second-order accurate fluxes. Both explicit and implicit time integration methods are available; in this work, rapid convergence to steady state is achieved with the data parallel line relaxation (DPLR) method~\cite{art:DPLR}.
US3D is equipped with chemistry/multi-temperature source terms and transport properties with the possibility to account for high temperature and high pressure effects.
Native routines can be further extended by user-defined subroutines, which allow coupling the solver to external libraries; we refer to Capriati et al.~\cite{capriati2021development} for the coupling with \mpp.

\subsection{Immersed Boundary Solvers}
\label{sec:INCA}

Two IB solvers are considered: one able to use both the cut-cell and the ghost-cell methods, and another using only the latter.

Employing a cut-cell IB methodology, INCA is a high-fidelity finite-volume solver for direct numerical simulations (DNS) and large eddy simulations (LES) of the compressible chemically reacting Navier-Stokes equations and provides a large number of different discretization schemes on three-dimensional block-Cartesian AMR grids~\cite{hickel2014subgrid,muller2016large}.
For the purposes of this study, a third-order weighted essentially non-oscillatory (WENO) scheme~\cite{jiang1996efficient} with HLLC flux function~\cite{toro2013riemann} is selected to discretize the inviscid terms.
WENO schemes permit high accuracy in smooth regions, while ensuring stable and sharp capturing of discontinuities. 
Second-order centered differences are used for the viscous terms and the explicit third-order Runge-Kutta scheme of Gottlieb and Shu~\cite{gottlieb1998total} is selected for time integration. 
Chemical source terms are treated using Strang's second-order time splitting scheme~\cite{strang1968construction} to alleviate the numerical stiffness caused by these terms. 
The chemical source terms thus reduce to a system of ordinary differential equations, which is solved by the VODE library \cite{brown1989vode}.
INCA employs a unique improvement to the common cut-cell methodology~\cite{meyer2010conservative}, which we refer to as the cut-element method~\cite{orley2015cut,pasquariello2016cut}.
This method represents the fluid-solid interface through cut-elements, which are derived from the Cartesian mesh and the triangulation of the surface geometry.
Instead of considering a planar intersection of a finite-volume cell with the wall surface, as typically done in cut-cell methods~\cite{ingram2003developments,schneiders2016efficient}, cut-elements maintain all details of the intersection of the grid with the surface triangulation.
The interface within each cut-cell is thus represented by several cut-elements belonging to different surface triangles to yield sub-cell accuracy and robustness for complex geometries.
This method is a consistent and conservative extension of the finite volume flux balance to accommodate cells being split by boundaries.
Further details on this cut\nobreakdash-element methodology and its extension to incorporate GSI and the effects of thermal nonequilibrium are provided in Ref.~\cite{baskaya3AF}.

INCA employs ghost-cells to allow for the use of unmodified stencils throughout the domain, as shown in~\autoref{fig:method_sketch_3}.
Moreover, the cut-cell procedure that ensures strict conservation can be switched off to use only the ghost-cell method. We will discuss results obtained with the INCA ghost-cell method for selected cases in~\autoref{sec:GP_vs_CC}. 

In contrast to INCA, the IB method implemented in the flow solver CHESS of Politecnico di Bari~\cite{ninni2022phd} fully relies on ghost-cells.
The numerical method utilized by CHESS is based on the flux vector splitting proposed by Steger and Warming~\cite{steger1981flux} with a second-order MUSCL reconstruction in space~\cite{van1979towards} for the hyperbolic terms. 
Discretization of the viscous fluxes uses Gauss's theorem in conjunction with a second-order linear reconstruction of the solution. 
A third-order explicit Runge-Kutta scheme is employed for time integration of the transport terms in the Navier-Stokes equations. 
Following the Runge-Kutta time step, chemical source terms are computed by means of a Gauss-Seidel scheme.
CHESS also uses AMR to provide appropriate resolution of shocks and boundary layers~\cite{de2007immersed2} and uses the same physicochemical models as US3D and INCA~\cite{baskaya2022verification}.
Further details on the solver can be found in the aforementioned works~\cite{baskaya2022verification,ninni2022phd}.

\section{Benchmark Cases} \label{sec:results}

We have curated a set of benchmark cases through collaborative effort with several research groups~\cite{baskaya2022verification}. 
The goal is to first verify the physicochemical models and the numerical schemes.
Once confidence is established over these fundamental aspects, the accuracy and limitations of the IB methods is addressed.
We have selected setups that are sufficiently challenging for the methods under assessment, and simple enough to be readily reproduced by others to incentivize collaboration.
For IB methods on Cartesian grids, curved geometries were selected to include the entire angular range of fluid-solid interfaces in two dimensions.
These cases include strong thermal gradients near cold isothermal walls as well as gas-surface interactions such as catalytic reactions and ablative surface blowing. 

The benchmark cases are summarized in~\autoref{tab:case_summary}. The first four cases serve as the verification of the implementations for chemistry, transport properties, the catalytic boundary conditions, and the numerical schemes for shock capturing.
Established validation experiments are chosen as the final benchmark cases: the fifth, sixth and seventh cases are 2-D cylinder flows of Knight et al.~\cite{knight2012assessment} with inert adiabatic, inert isothermal, and catalytic isothermal surfaces. As the eighth benchmark, we discuss results for an ablative TPS sample geometry under plasma wind tunnel conditions, for which reference experimental data is provided by Helber et al.~\cite{helber2017determination}.

\begin{table}[h]
\footnotesize
	\centering
	\caption{Summary of studied cases.}
	\label{tab:case_summary}
	\begin{tabular}{@{}llll@{}}
		\toprule
		 & Name & Aspect to Assess & Section \\ \midrule
	    1. & 0-D Reactor & Chemistry & \ref{sec:0D_reactor}       \\
		2. & 1-D Diffusion Problem & Mass diffusion & \ref{sec:1D_boundary}       \\
		3. & 1-D Catalytic Diffusion Problem & Mass diffusion with catalysis & \ref{sec:cata1D_boundary}        \\
		4.  & 1-D Shocktube & Shock capturing & \ref{sec:shocktube}       \\
		5.  & 2-D Cylinder (inert, adiabatic wall) &  Chemical nonequilibrium & \ref{sec:knight_adia}      \\
		6.  & 2-D Cylinder (inert, isothermal wall) & Surface heat flux & \ref{sec:knight_isot}       \\
		7.  & 2-D Cylinder (fully catalytic, isot. wall) & Surface heat flux with catalysis & \ref{sec:knight_cata}       \\
		8.  & 2-D Ablator & Surface mass blowing with ablation & \ref{sec:abla}       \\ \bottomrule
	\end{tabular}
\end{table}

\subsection{0-D Reactor} \label{sec:0D_reactor}

\begin{figure}[tb]
    \centering
    \includegraphics[clip, trim=0cm 0cm 0cm 0cm, width=.499\textwidth]{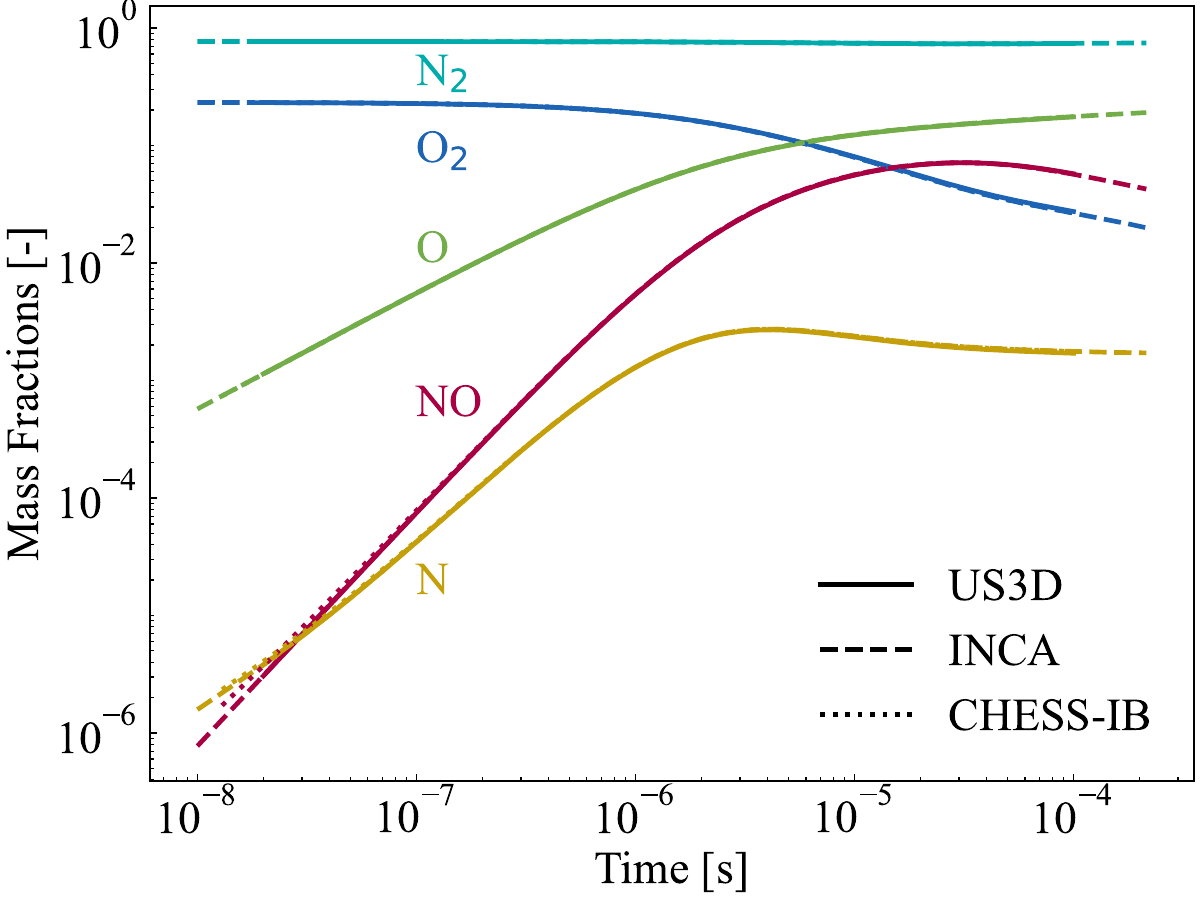}
    \caption{Evolution of mass fractions for 5-species air in the 0-D reactor case.}
    \label{fig:chem_box_results_air5}
\end{figure}

The first study verifies the chemical source term implementation by considering 5-species air, $ [ \text{N}_2, \text{O}_2, \text{NO}, \text{N}, \text{O} ] $, in an adiabatic reactor. Starting from the chemical nonequilibrium (CNE) initialization provided in \autoref{tab:0D_reactor}, the system is left free to time-march towards the equilibrium state according to chemical mechanisms from Park~\cite{park1993review,park2001chemical}.  
The solutions provided by all three solvers are shown in~\autoref{fig:chem_box_results_air5}. 
Dissociation of $ \text{N}_2, \text{O}_2 $ and the resulting formation of $ \text{NO}, \text{N}, $ and $ \text{O} $ can be seen.
The code-to-code agreement is excellent. 

\begin{table}[h]
	\centering
	\caption{Setup conditions for the 0-D reactor case.}
	\label{tab:0D_reactor}
	\begin{tabular}{@{}ccccc@{}}
		\toprule
		$ \rho $ [kg/m$ ^3 $] & $ T $ [K]  & $ u $ [m/s] &  $ y(\text{N}_2) $ & $ y(\text{O}_2$)   \\ \midrule
		0.01 & 7000 & 0.0 &  0.767  &  0.233  \\ \bottomrule
	\end{tabular}
\end{table}

\subsection{1-D Diffusion Problem} \label{sec:1D_boundary}

This test case verifies the implementation of models for transport properties. Viscosity and thermal conductivity are obtained through direct calls to \mpp, and are exactly equal for all solvers.
Therefore, mainly the differences in the implementation of the driving force and boundary conditions are assessed. 
The setup consists of a 1-D tube with isothermal end walls at different temperatures. The initial and boundary conditions are provided in \autoref{tab:1D_boundary}. 
The mixture composition and reaction mechanisms are the same as in the 0-D reactor case. The tube is \SI{3}{\milli\meter} long.
It should be pointed out that the computational meshes in US3D and INCA solvers have 100 cells, whereas CHESS results~\cite{baskaya2022verification} used 400 cells. It has been verified that the US3D and INCA solutions are grid converged on the mesh with 100 cells.

\begin{figure} [tb]
	\begin{subfigure}[t]{0.49\textwidth}\centering
		\includegraphics[clip, trim=0cm 0cm 0cm 0cm, 
		width=0.999\columnwidth]{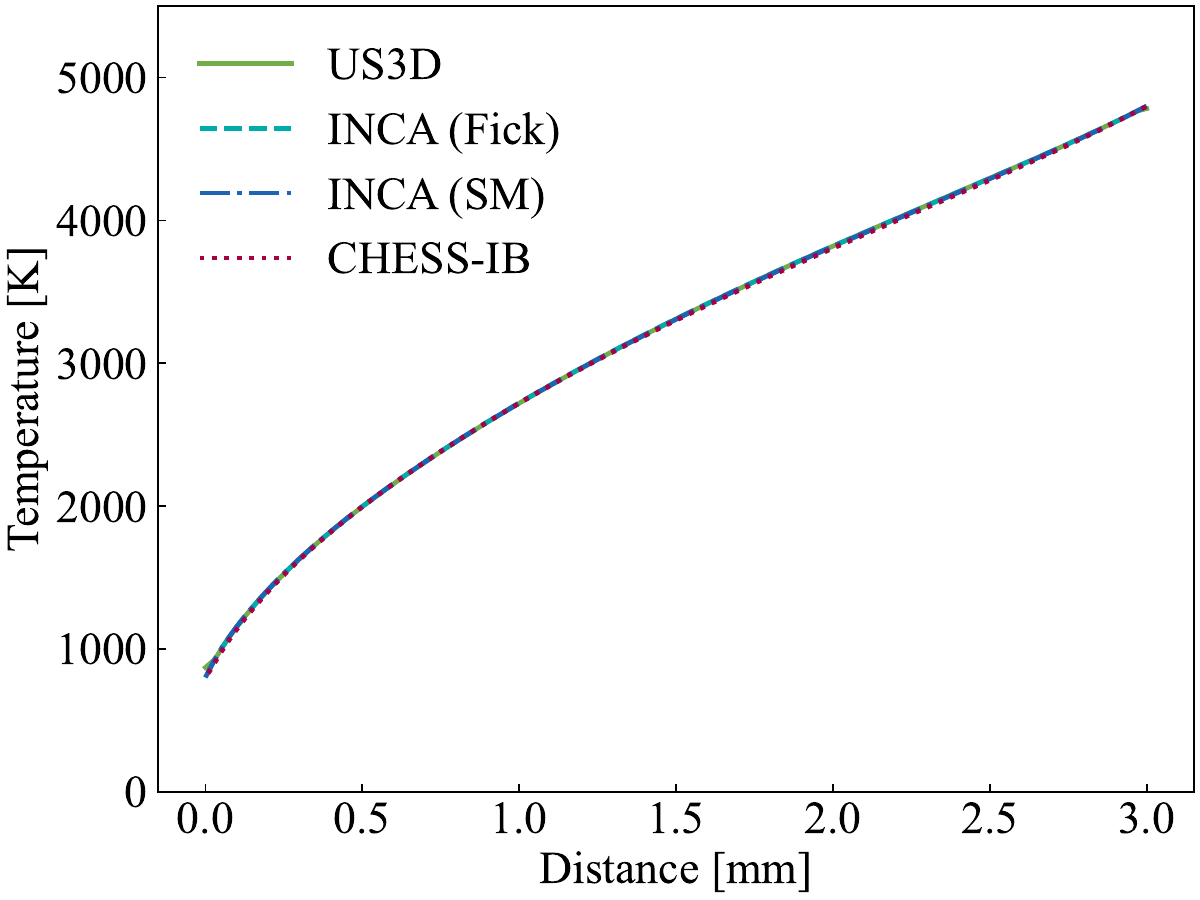}
		\caption{}
	\end{subfigure}
    \hspace{0.1cm}
	\begin{subfigure}[t]{.49\textwidth}\centering
		\includegraphics[clip, trim=0cm 0cm 0cm 0cm, 
		width=0.999\columnwidth]{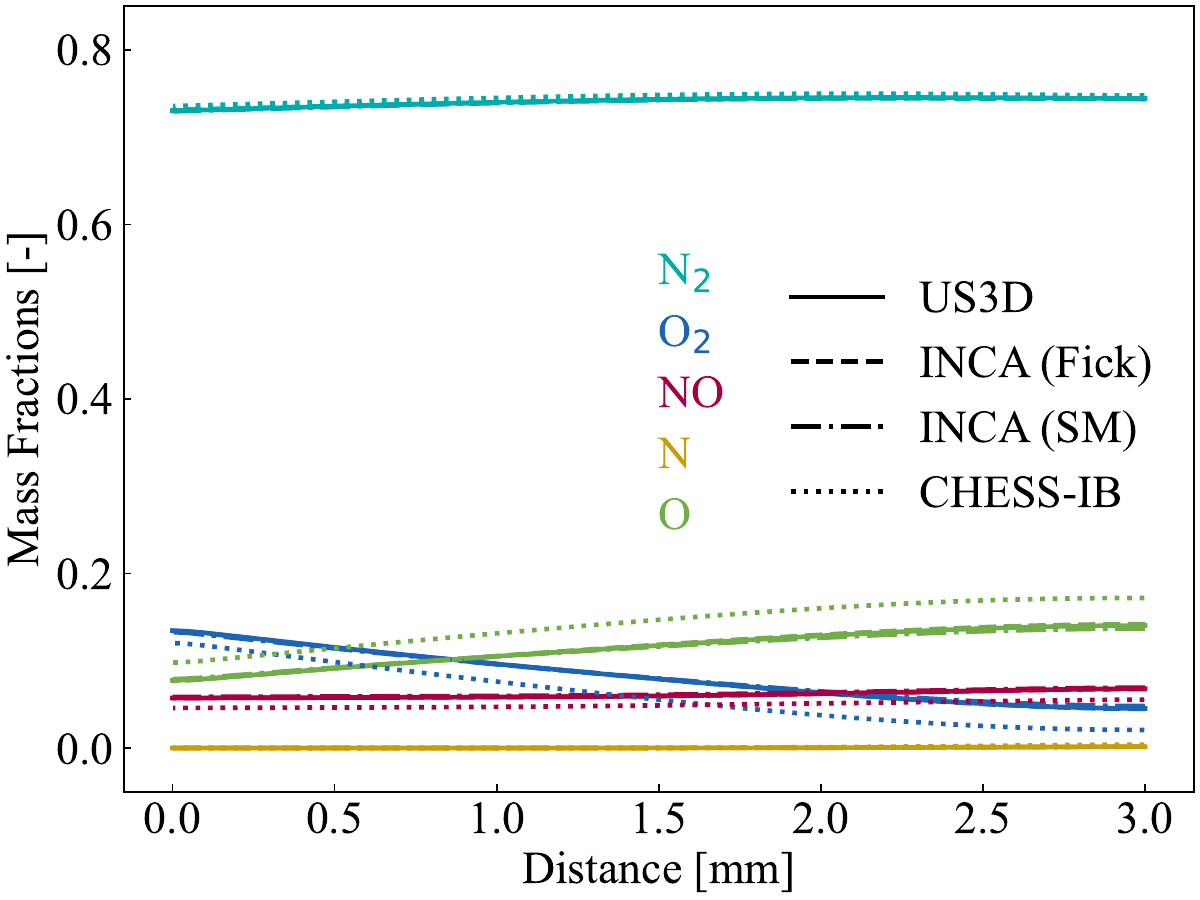}
		\caption{}
	\end{subfigure}
	\caption{Comparison of (a) temperature and (b) mass fraction distributions for the 1-D diffusion case.}
	\label{fig:bound_case}
\end{figure}

\begin{table}[h]
	\centering
	\caption{Setup conditions for the 1-D diffusion case.}
	\label{tab:1D_boundary}
	\begin{tabular}{@{}ccccccc@{}}
		\toprule
		$ \rho $ [kg/m$ ^3 $] & $ T $ [K]  & $ T_{\text{left}} $ [K]  & $ T_{\text{right}} $ [K]  & $ u $ [m/s] &  $ y(\text{N}_2) $ & $ y(\text{O}_2) $  \\ \midrule
		0.02 & 1000 & 800 & 4800 & 0.0 &  0.767  &  0.233  \\ \bottomrule
	\end{tabular}
\end{table}
In this test case the temperature gradient leads to chemical reactions, which in turn drive mass diffusion.
Temperature and mass fraction distributions along the tube are presented in~\autoref{fig:bound_case}. INCA results have been obtained by both Fick's law and Stefan-Maxwell diffusion models. However, for this test case, differences seem to be negligible between the two. Overall, US3D results are matched perfectly with INCA, while slight differences are observed for the mass fraction distributions predicted by CHESS, even though the temperature profiles match exactly. 

\subsection{1-D Catalytic Diffusion Problem} \label{sec:cata1D_boundary}

This test case verifies the catalytic boundary condition implementation for a simple [N$_2$, N] binary mixture along a 1-D tube, for which an analytical solution exists and is derived in the~\hyperlink{sec:appendix}{appendix}. 
Setup conditions are given in~\autoref{tab:1D_boundary_cata}. The length of the tube is \SI{0.2}{\meter}. One side of the tube at $\text{x}=0.0$~m is at reservoir conditions, while at the other, at $\text{x}=0.2$~m, a catalytic wall boundary condition is imposed. The catalytic wall promotes the recombination of nitrogen, that is, $ \text{N} + \text{N} \rightarrow \text{N}_2 $. The reaction rate is controlled by the recombination coefficient $\gamma$ through~\autoref{eq:react_gamma}.

\begin{table}[h]
	\centering
	\caption{Setup conditions for the 1-D catalytic diffusion case.}
	\label{tab:1D_boundary_cata}
	\begin{tabular}{@{}ccccccc@{}}
		\toprule
		$ p $ [Pa] & $ T $ [K]  & $ T_{\text{wall}} $ [K] & $ u $ [m/s] &  $ y(\text{N}_2) $ & $ y(\text{N}) $ & $ \gamma_\text{N} $  \\ \midrule
		100 & 3000 & 3000 & 0.0 &  0.0  &  1.0 & [0.001, 0.01, 0.1, 1.0] \\ \bottomrule
	\end{tabular}
\end{table}

\begin{figure}[tb]
    \centering
    \includegraphics[clip, trim=0cm 0cm 0cm 0cm, width=.499\textwidth]{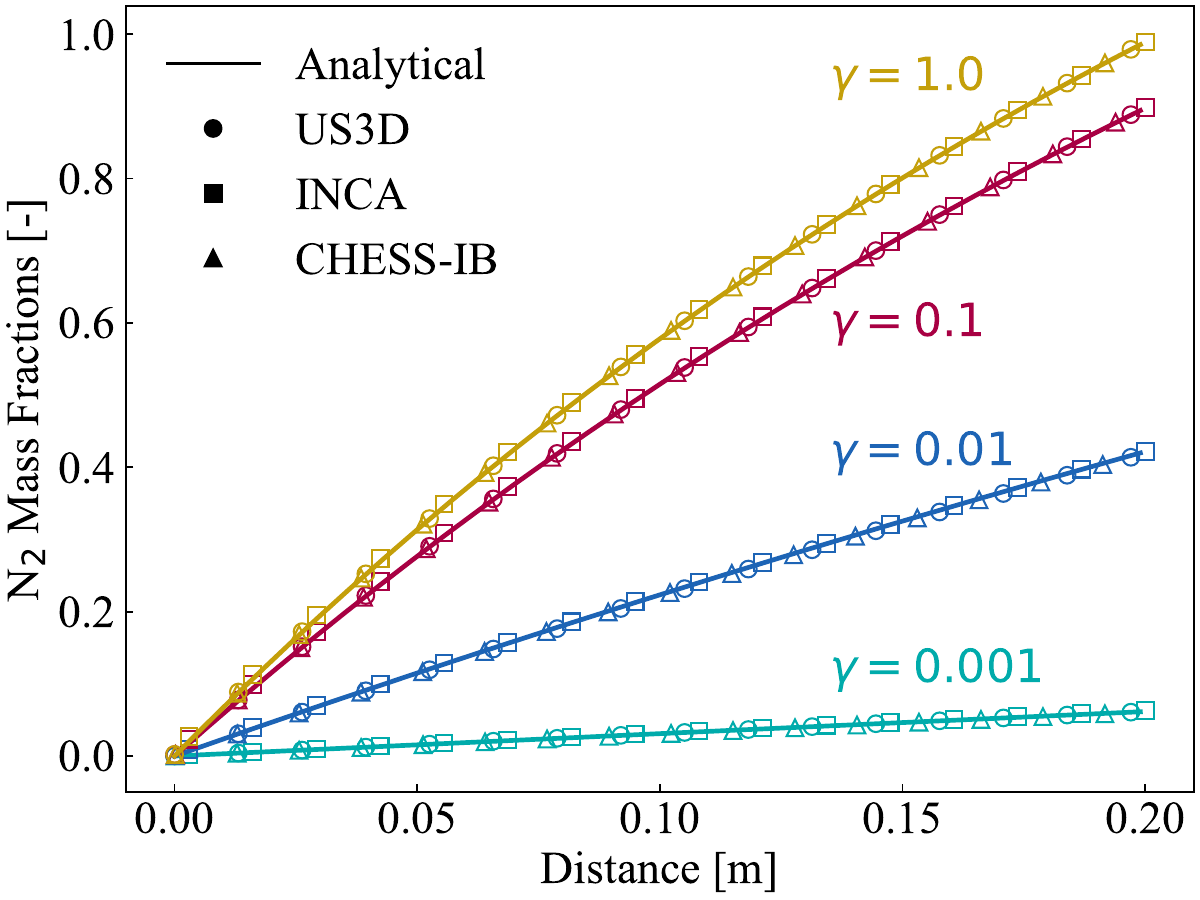}
    \caption{$\text{N}_2$ mass fractions for different recombination coefficients $\gamma$ for the 1-D catalytic diffusion problem.}
    \label{fig:Ycata_bound}
\end{figure}

Results obtained with US3D, INCA, and CHESS~\cite{baskaya2022verification} are compared with the analytical reference solution in~\autoref{fig:Ycata_bound}. Naturally, for higher values of the recombination coefficient $\gamma$, mass fraction of molecular nitrogen at the wall increases, and reaches unity for the fully catalytic case with $\gamma = 1.0$.
All numerical predictions are in excellent agreement with the analytical solution.
The previously noted difference for the CHESS solver in the diffusion problem is not observed here as the diffusion of species are driven predominantly by the surface reactions.

\subsection{1-D Shocktube} \label{sec:shocktube}

The Riemann problem of Grossman	and Cinnella~\cite{grossman1990flux} is used to evaluate shock capturing. The unit domain, $\text{x}=[0,1]$~m, is spatially discretized by 600 cells, in line with the reference resolution. 
The diaphragm separating the two initial states is set at the midpoint of the tube. The initial conditions for the two states are given in \autoref{tab:shocktube}. 
Air with 5-species is initially considered at thermodynamic equilibrium.  
The reaction mechanism is taken from an earlier work of Park~\cite{park1985convergence}, to match with the reference~\cite{grossman1990flux}. 
Grossmann and Cinnella applied a thermal nonequilibrium model; however, we have performed tests with Park’s two-temperature model~\cite{park1989nonequilibrium} and found no significant differences between the translational and vibrational energy modes. Therefore, we show results that have been obtained with a thermal equilibrium assumption.

\begin{table}[h]
	\centering
	\caption{Initial conditions for the 1-D shocktube case.}
	\label{tab:shocktube}
	\begin{tabular}{@{}ccccccc@{}}
		\toprule
		$ u_{left} $ [m/s] & $ T_{left} $ [K]  & $ p_{left} $ [Pa]  & $ u_{right} $ [m/s] & $ T_{right} $ [K]  & $ p_{right} $ [Pa]  \\ \midrule
		0.0 & 9000 & 195256 & 0.0 & 300 &  10000  \\ \bottomrule
	\end{tabular}
\end{table}

\begin{figure} [tb]
	\begin{subfigure}[t]{0.49\textwidth}\centering
		\includegraphics[clip, trim=0cm 0cm 0cm 0cm, 
		width=0.999\columnwidth]{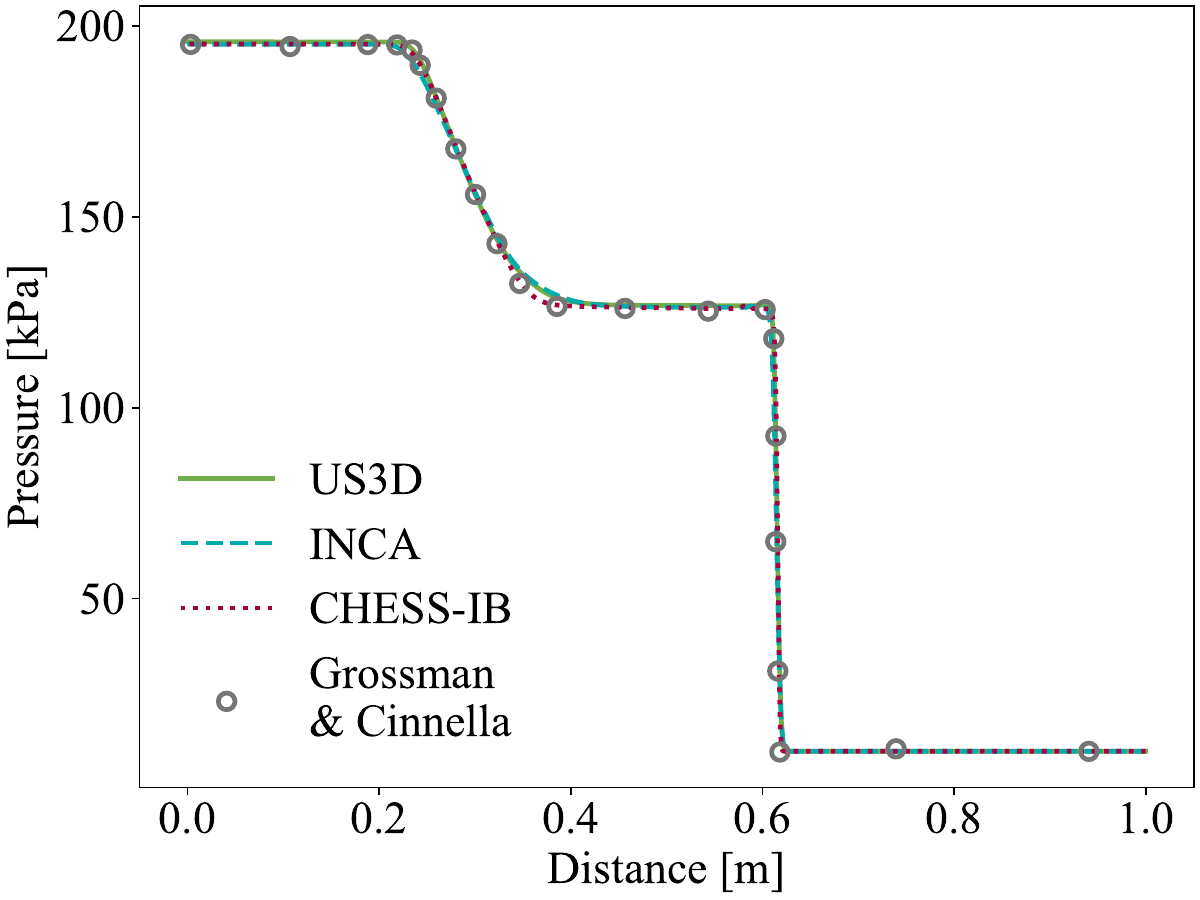}
		\caption{}
	\end{subfigure}
 \hspace{0.1cm}
	\begin{subfigure}[t]{0.49\textwidth}\centering
		\includegraphics[clip, trim=0cm 0cm 0cm 0cm, 
		width=0.999\columnwidth]{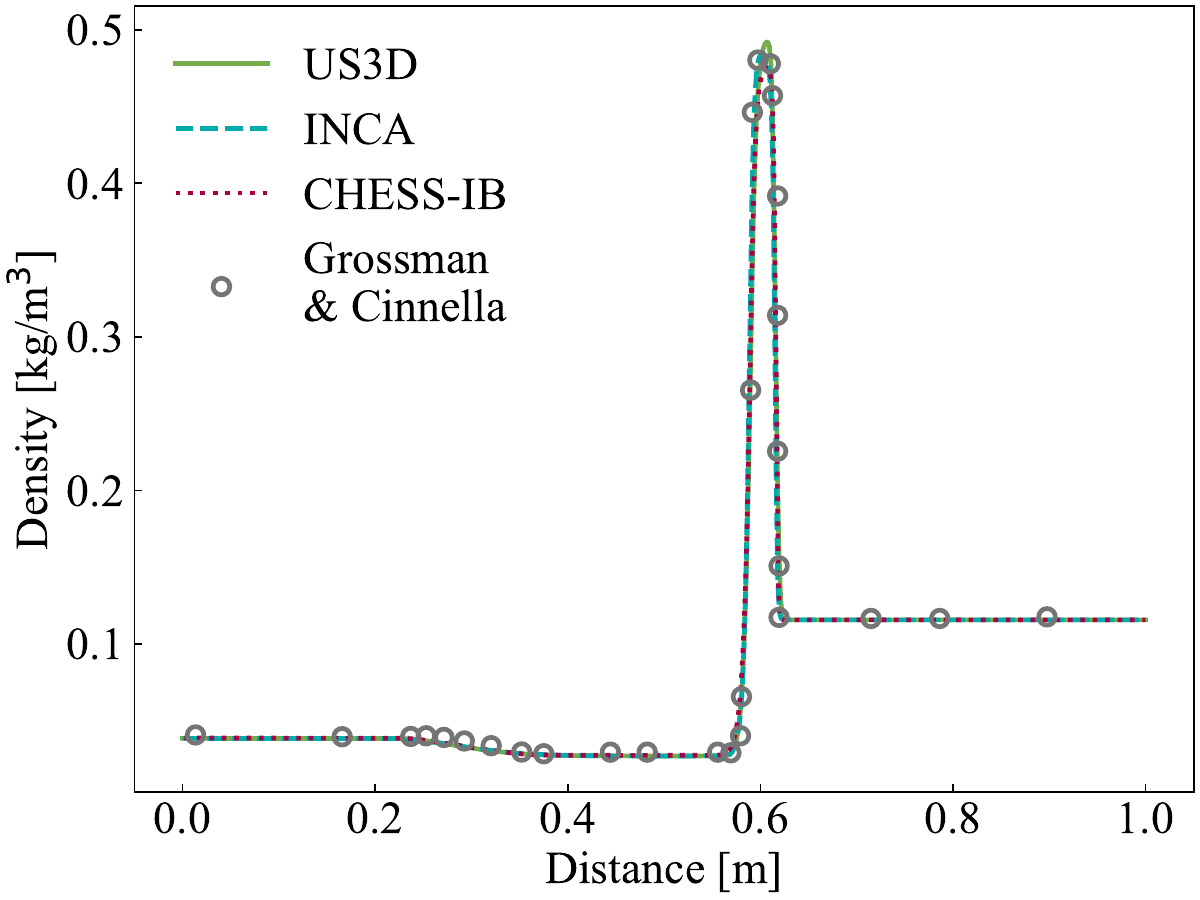}
		\caption{}
	\end{subfigure}
	\caption{Comparison of (a) pressure and (b) density distributions for the 1-D shocktube case of Grossman and Cinnella~\cite{grossman1990flux}.}
	\label{fig:shocktube}
\end{figure}

\begin{figure} [tb]
    \begin{subfigure}[t]{0.49\textwidth}\centering
		\includegraphics[clip, trim=0cm 0cm 0cm 0cm, 
		width=0.999\columnwidth] {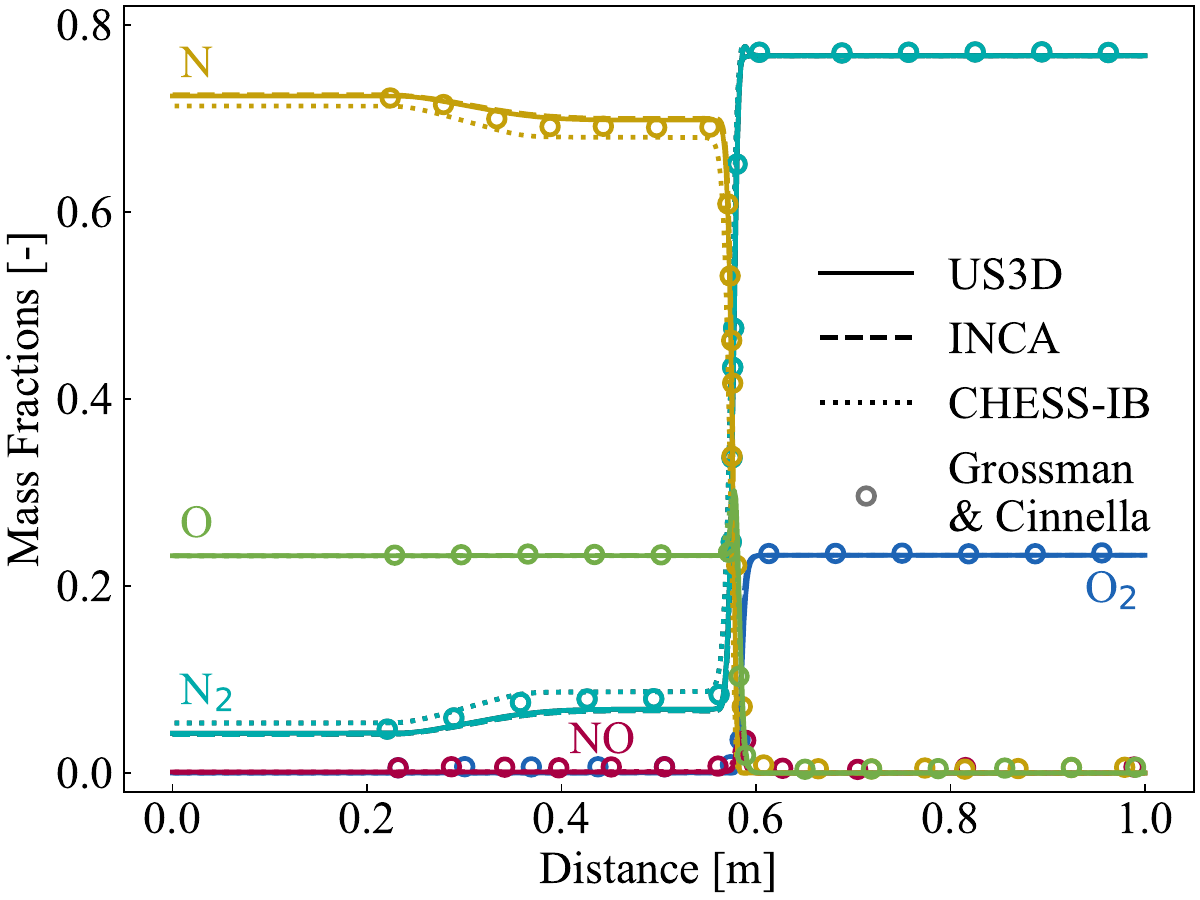}
		\caption{}
        \label{fig:shocktube_mass_a}
	\end{subfigure}
 \hspace{0.1cm}
	\begin{subfigure}[t]{0.49\textwidth}\centering
		\includegraphics[clip, trim=0cm 0cm 0cm 0cm, 
		width=0.999\columnwidth] {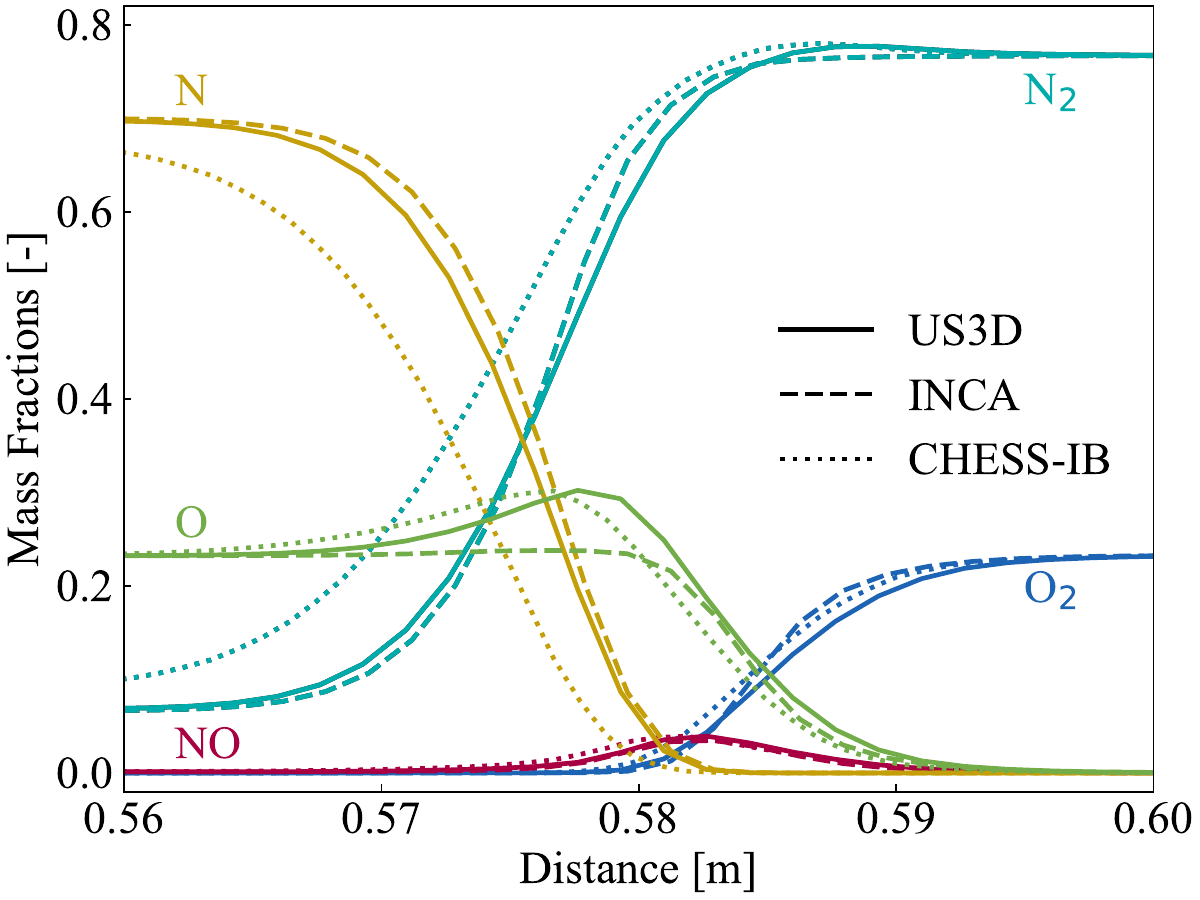}
		\caption{}
  \label{fig:shocktube_mass_b}
	\end{subfigure}
	\caption{Comparison of mass fraction distributions for the 1-D shocktube case.}
    \label{fig:shocktube_mass}
\end{figure}

\autoref{fig:shocktube} shows pressure and density profiles \SI{99}{\micro\second} after the initial state. Mass fractions are given in~\autoref{fig:shocktube_mass}. 
The contact discontinuity and the shock wave traveling in the positive x direction as well as the expansion traveling in the opposite direction are well captured. The peak in density after the shock also matches perfectly with the reference results without any oscillations.
Predictions of US3D and INCA for the mass fractions are also in excellent agreement with the reference results of Grossmann and Cinnella, see~\autoref{fig:shocktube_mass_a}. The minor differences between the solvers in their sharp representation of the discontinuity is shown in the close-up view in~\autoref{fig:shocktube_mass_b}. 
CHESS~\cite{baskaya2022verification} predicts a slightly higher $\text{N}_2$ mass fraction, and accordingly less atomic nitrogen, than US3D and INCA.

\subsection{2-D Cylinder} \label{sec:knight}

2-D cylinder flows are used for the validation of surface heat flux calculations under inert and catalytic wall conditions.
Knight et al.~\cite{knight2012assessment} have presented an assessment of five different CFD codes from participating institutions with respect to reference experiments conducted at the high\nobreakdash-enthalpy shock tunnel of the German Aerospace Center (DLR)~\cite{karl2003high}. The experiment investigates the flow past a cylinder with a radius of \SI{0.045}{\meter} exposed to a reported total enthalpy of \SI{22.4}{\mega\joule/\kilogram}.
The experimental setup is numerically replicated by imposing the inflow conditions given in \autoref{tab:knight} on the left boundary. Symmetry is imposed along the stagnation line, and the outer boundaries are set as non\nobreakdash-reflecting outlets. The reaction mechanism employed for the 5-species air model is taken from Park~\cite{park1993review,park2001chemical}. 
As remarked by Knight et al.~\cite{knight2012assessment}, there appears to be a large variation in the results from different solvers, especially regarding the treatment of the surface. 
To study this sensitivity, three different surface conditions are tested in the following sections: two inert cases with adiabatic and isothermal conditions, and a third case with a fully catalytic isothermal wall.

In the following assessment of IB methods, ``BF'' is used to denote reference results obtained on body-fitted grids with US3D, ``IB-CC'' is used for the cut-cell IB method of INCA, and ``IB-GP'' refers to the ghost-cell IB method of CHESS~\cite{baskaya2022verification}.

\begin{table}[h]
	\centering
	\caption{Freestream conditions for the 2-D cylinder case.}
	\label{tab:knight}
	\begin{tabular}{@{}ccccc@{}}
		\toprule
		$ \text{M}_\infty $ & $ u_\infty $ [m/s] & $ T_\infty $ [K] & $ p_\infty $ [Pa]  & $ \rho_\infty $ [kg/m$ ^3 $] \\ \midrule
            8.98 & 5956 & 901 & 476 & 1.547$ \times10^{-3} $ \\
            \midrule
            \midrule
  $ y(\text{N}_2) $ & $ y(\text{O}_2$) &  $ y(\text{NO}) $ & $ y(\text{N}$) &  $ y(\text{O}) $  \\ \midrule
		0.7543 & 0.00713 & 0.01026 & $ 6.5\times10^{-7} $ & 0.2283  \\ \bottomrule
	\end{tabular}
\end{table}

\subsubsection{Inert Adiabatic Wall} \label{sec:knight_adia}

\begin{figure} [tb]
	\begin{subfigure}[t]{.49\textwidth}\centering
		\includegraphics[clip, trim=0cm 0cm 0cm 0cm, 
		width=0.999\columnwidth]{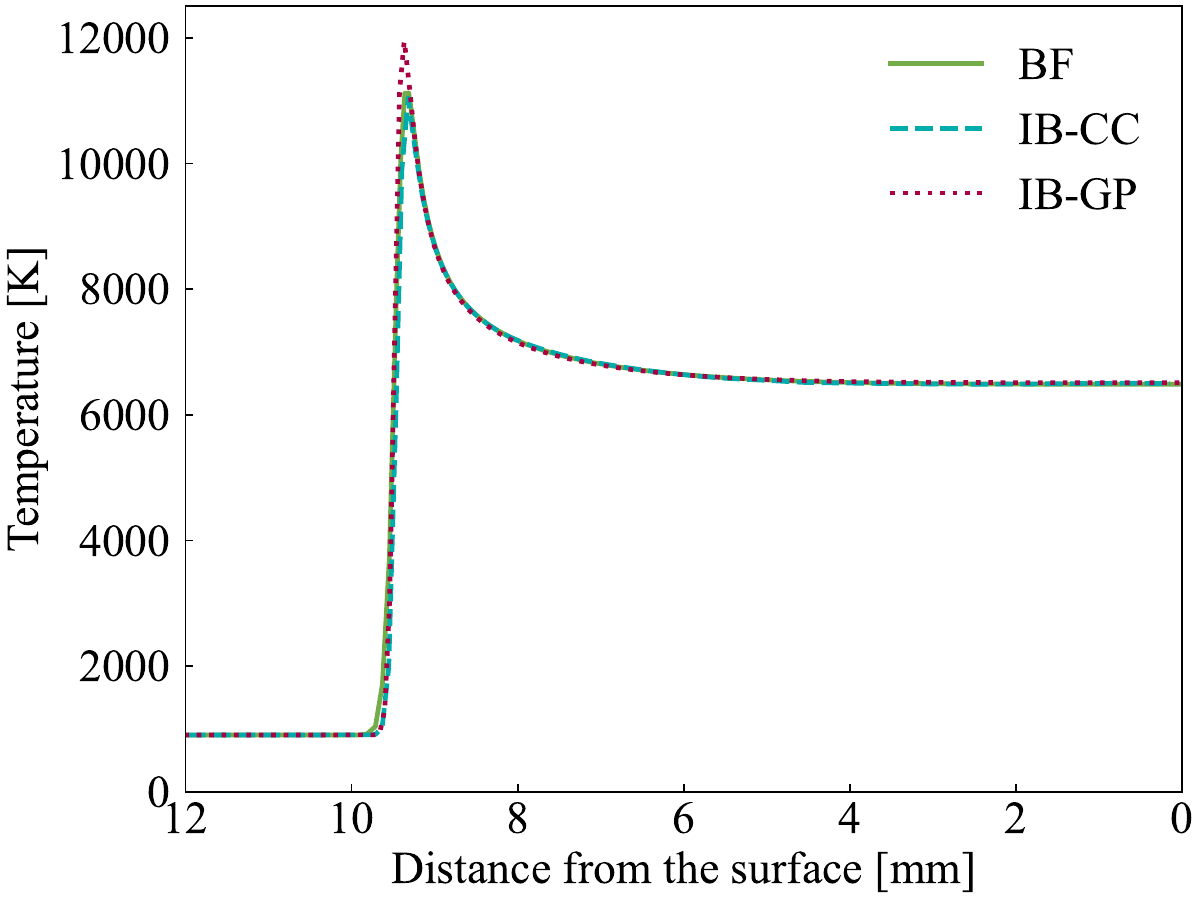}
		\caption{}
	\end{subfigure}
 \hspace{0.1cm}
	\begin{subfigure}[t]{.49\textwidth}\centering
		\includegraphics[clip, trim=0cm 0cm 0cm 0cm, 
		width=0.999\columnwidth]{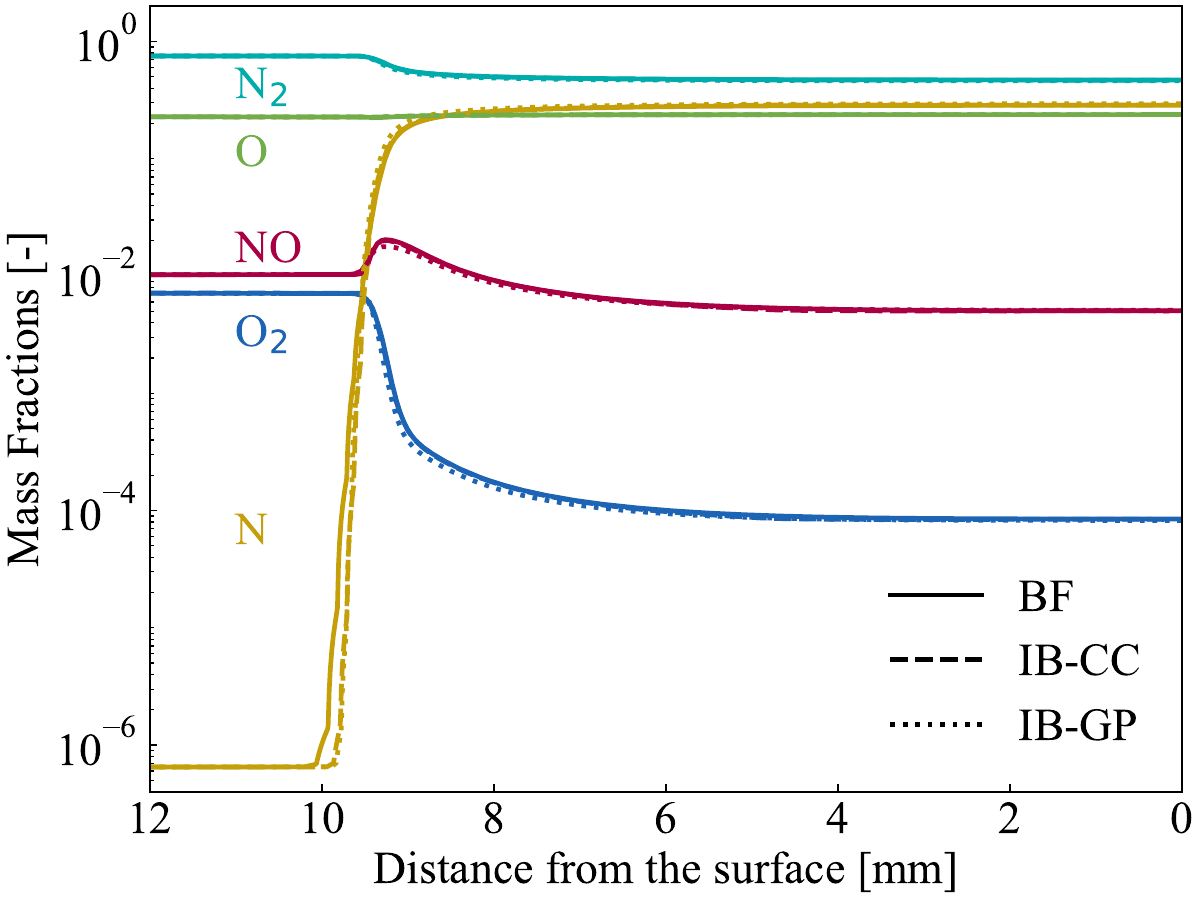}
		\caption{}
	\end{subfigure}
	\caption{Comparison of (a) temperature and (b) mass fractions along the stagnation line for the inert adiabatic 2-D cylinder case. }
	\label{fig:knight_inert_adia_stag}
\end{figure}

The temperature and species mass fractions along the stagnation line are presented for the adiabatic case in \autoref{fig:knight_inert_adia_stag}.
Shock stand-off distance and the dissociation of molecular nitrogen and oxygen in the shock layer are predicted in very good agreement between all methods.
The fundamental differences in the implementation of the adiabatic wall boundary condition have no noticeable effect on the results. This is in line with the expectation that truncation and conservation errors are small in the absence of strong gradients.

\begin{figure} [!h]
	\begin{subfigure}[t]{.49\textwidth}\centering
		\includegraphics[clip, trim=0cm 0cm 0cm 0cm, 
		width=0.999\columnwidth]{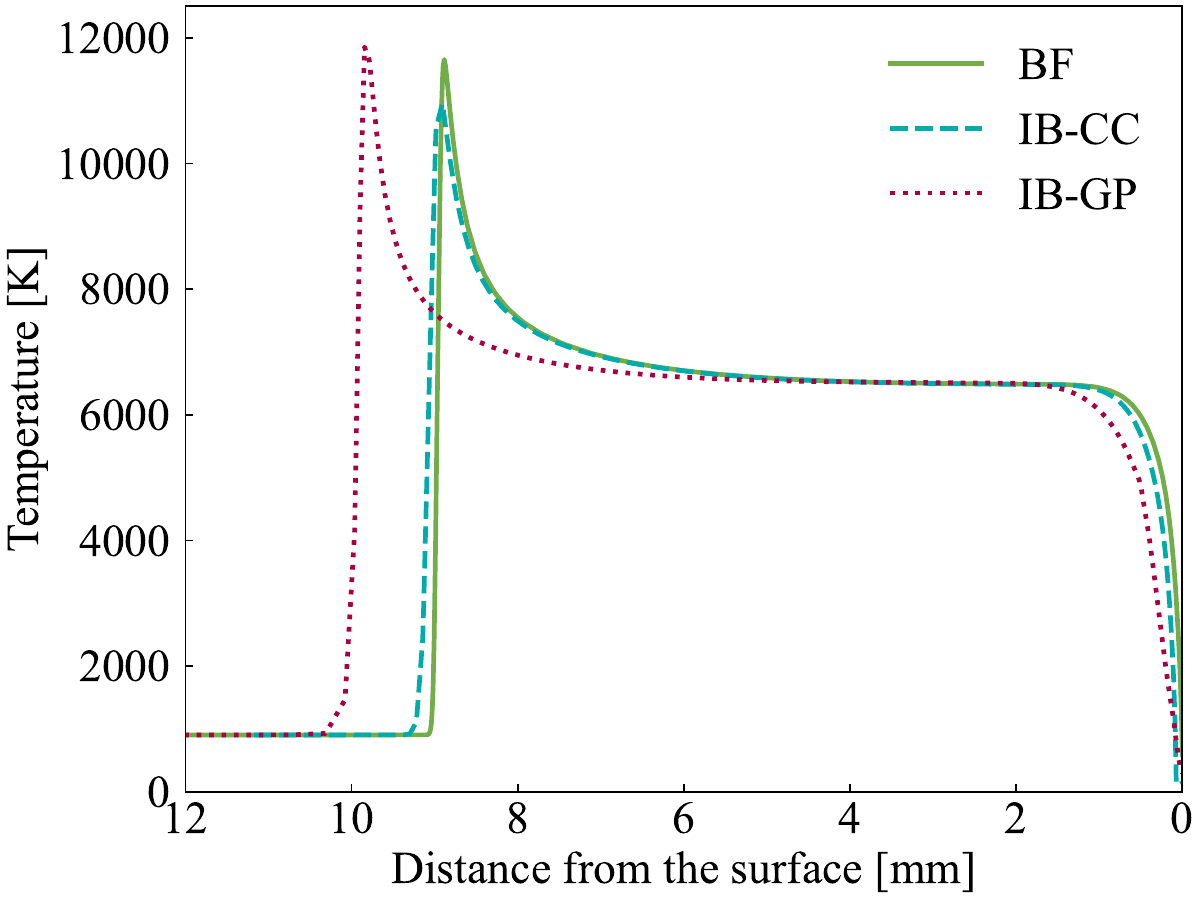}
		\caption{}
	\end{subfigure}
 \hspace{0.1cm}
	\begin{subfigure}[t]{.49\textwidth}\centering
		\includegraphics[clip, trim=0cm 0cm 0cm 0cm, 
		width=0.999\columnwidth]{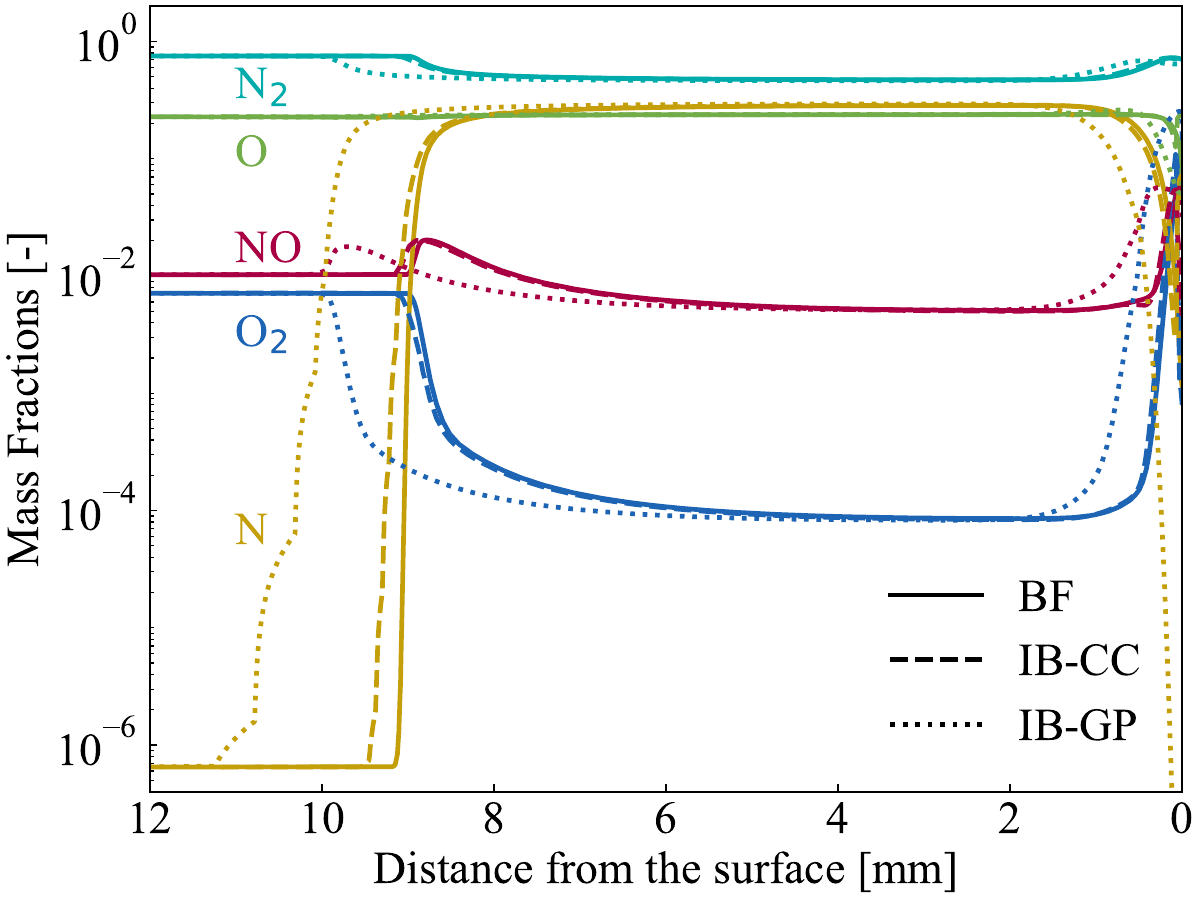}
		\caption{}
	\end{subfigure}
	\caption{Comparison of (a) temperature and (b) mass fractions along the stagnation line for the inert isothermal 2-D cylinder case by Knight et al.~\cite{knight2012assessment}.}
	\label{fig:knight_inert_iso_stag}
\end{figure}

\subsubsection{Inert Isothermal Wall} \label{sec:knight_isot}

For the same inflow conditions, an isothermal wall boundary condition with a wall temperature of \SI{300}{\kelvin} is imposed on the cylinder surface, in accordance with the specifications by Knight et al.~\cite{knight2012assessment}.
The numerical predictions for the stagnation line temperature and mass fraction distributions are plotted in~\autoref{fig:knight_inert_iso_stag}. 
Results obtained with the BF and the IB-CC methods match almost exactly, including the steep temperature and species variations in the boundary layer.
Results obtained with the IB-GP method, on the other hand, show a significant difference in the shock stand-off distance.
This could be attributed to the non-conservative formulation of the ghost-cell IB methodology. Mass conservation errors could manifest as an unphysical blowing from the surface. Consequently, the shock stand-off distance is increased and the whole flow field is modified.
The adiabatic case is less affected by these conservation errors because it has much smaller temperature and density gradients near the wall. 
The IB-CC method handles large temperature and density gradients at isothermal walls much better, because it uses a strictly conservative IB method. 

\begin{figure} [tb]
	\begin{subfigure}[t]{.49\textwidth}\centering
		\includegraphics[clip, trim=0cm 0cm 0cm 0cm, 
		width=0.999\columnwidth]{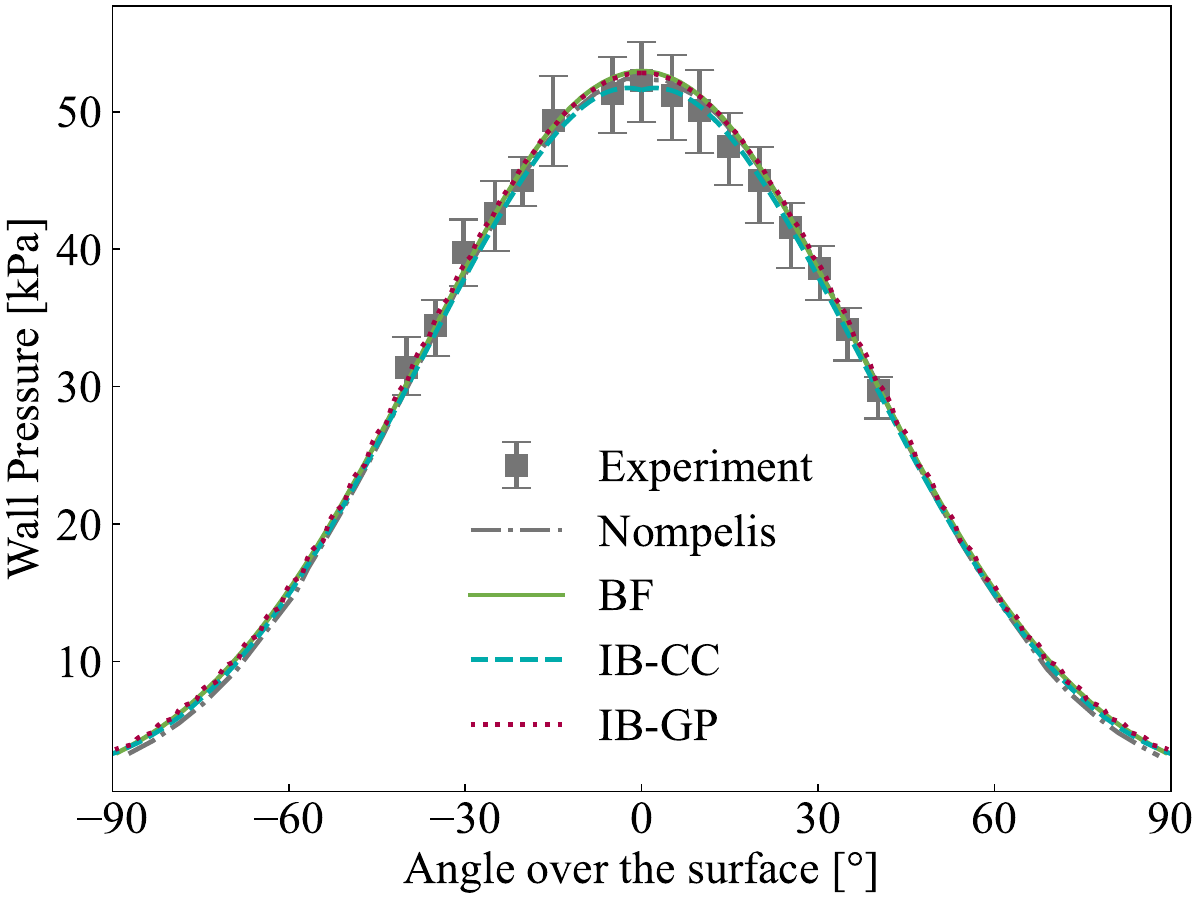}
		\caption{}
	\end{subfigure}
 \hspace{0.1cm}
	\begin{subfigure}[t]{.49\textwidth}\centering
		\includegraphics[clip, trim=0cm 0cm 0cm 0cm, 
		width=0.999\columnwidth]{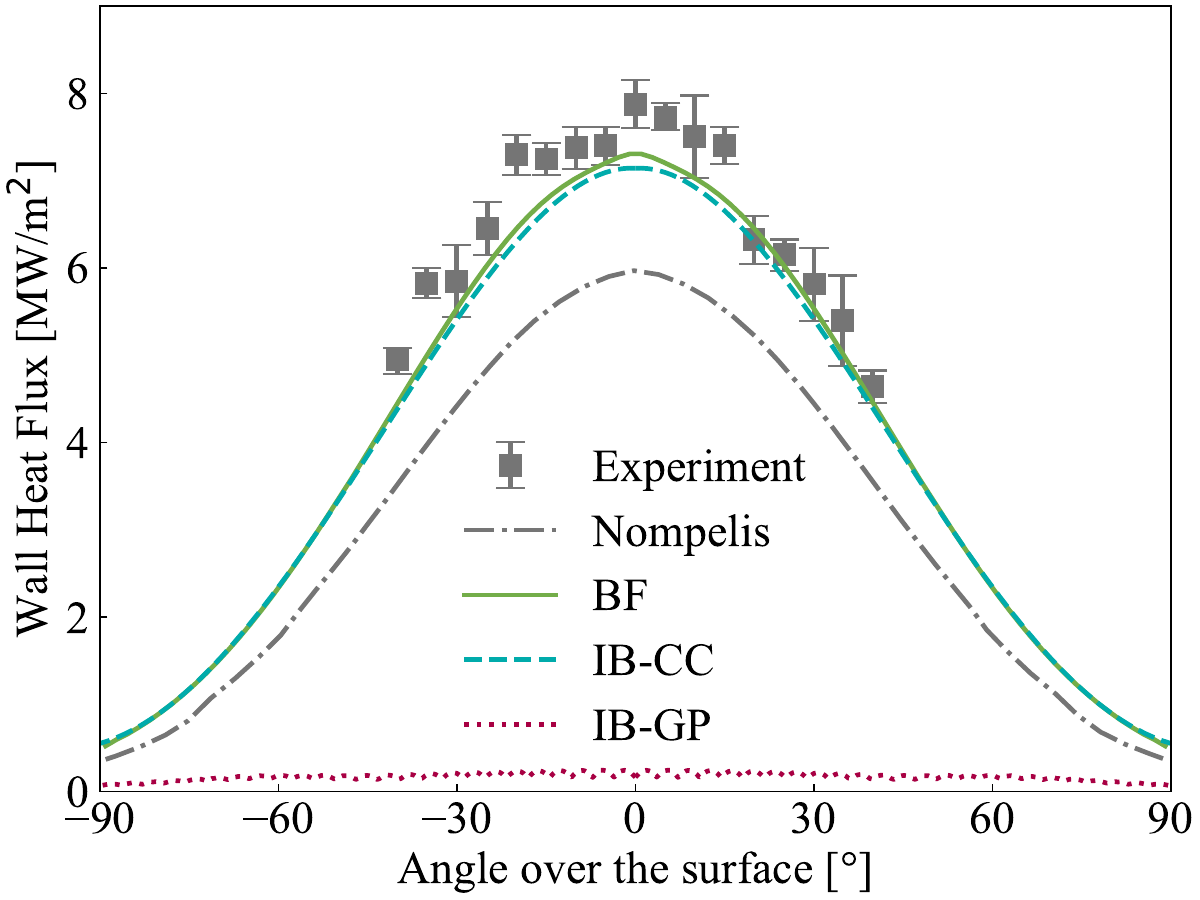}
		\caption{}
	\end{subfigure}
	\caption{Comparison of surface (a) pressures and (b) heat fluxes for the inert isothermal 2-D cylinder case by Knight et al.~\cite{knight2012assessment}.}
\label{fig:knight_inert_iso_heat_pres}
\end{figure}

\begin{figure} [tb]
	\begin{subfigure}[t]{.49\textwidth}\centering
		\includegraphics[clip, trim=0cm 0cm 0cm 0cm, 
		width=0.999\columnwidth]{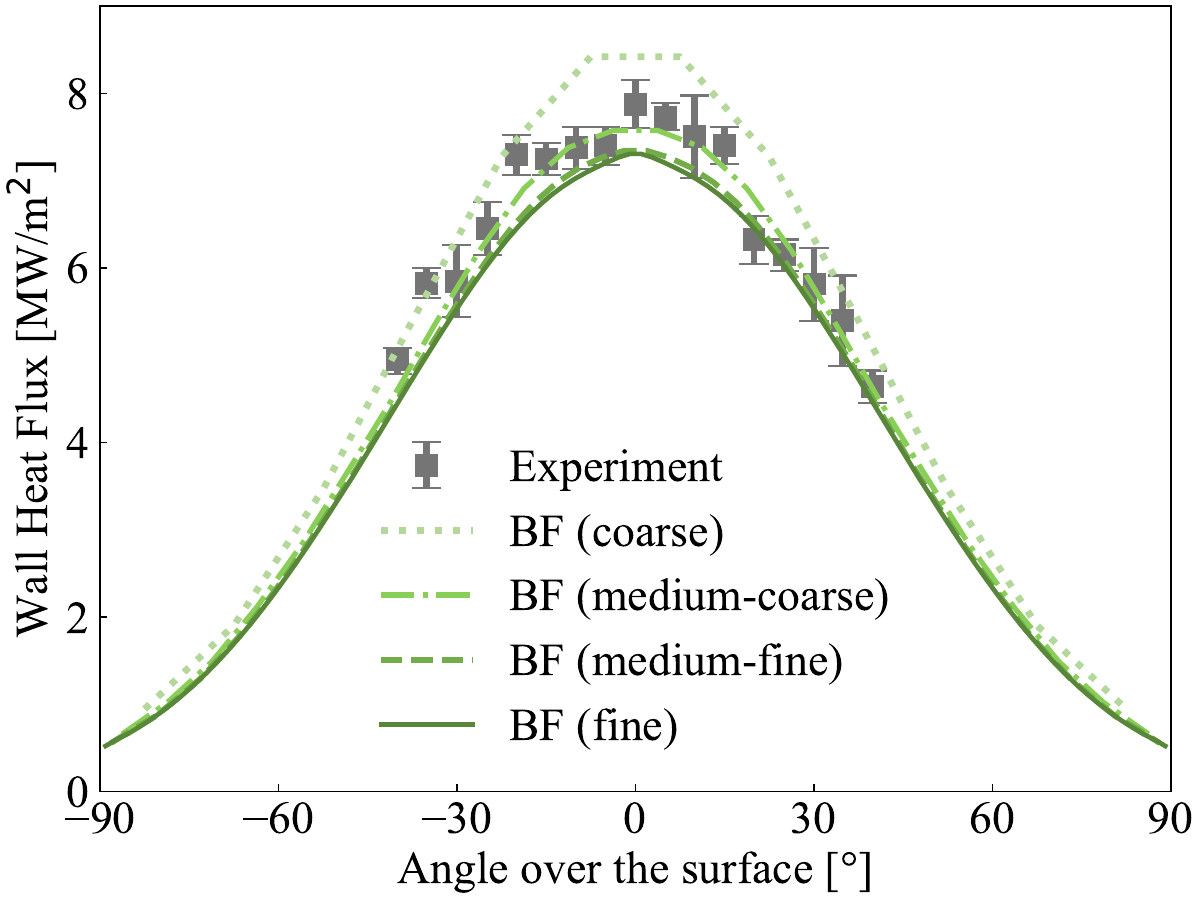}
		\caption{}
	\end{subfigure}
 \hspace{0.1cm}
	\begin{subfigure}[t]{.49\textwidth}\centering
		\includegraphics[clip, trim=0cm 0cm 0cm 0cm, 
		width=0.999\columnwidth]{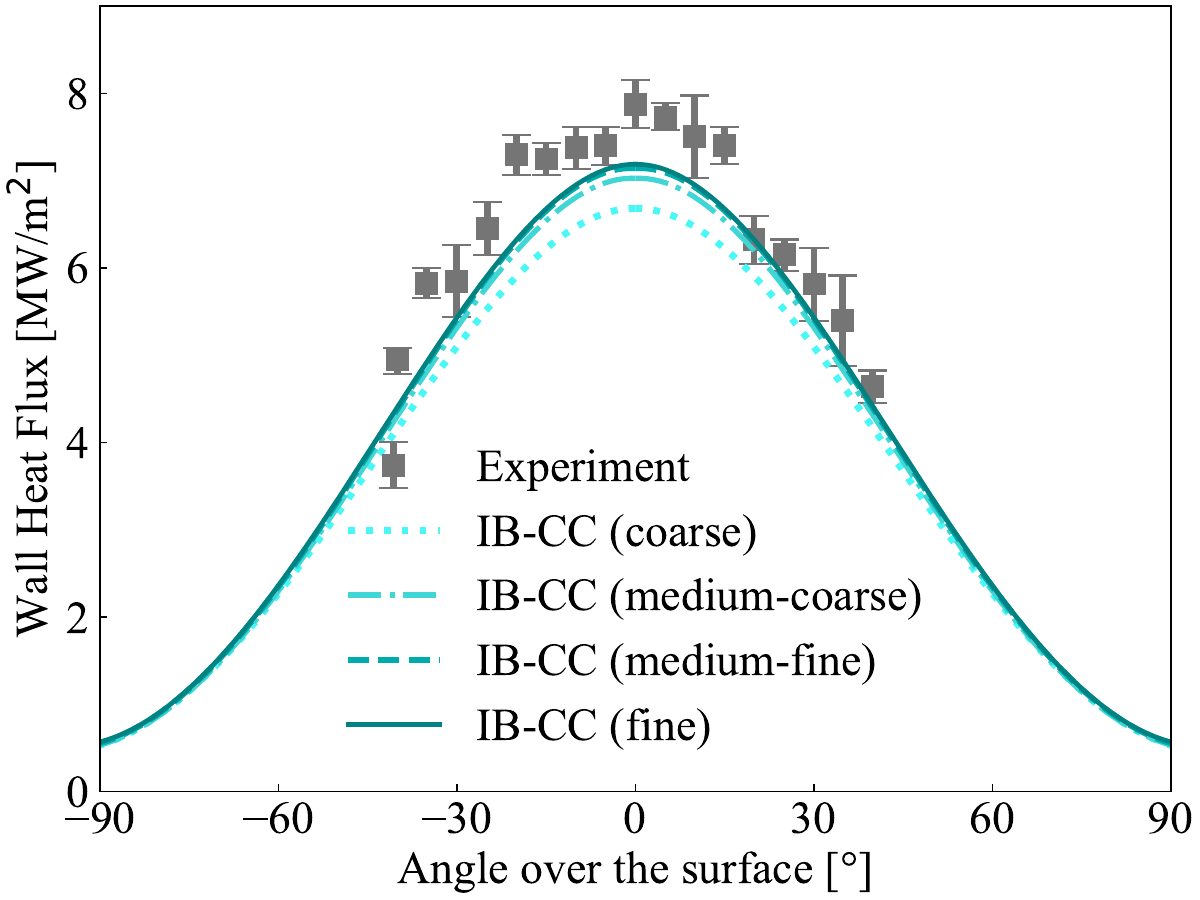}
		\caption{}
	\end{subfigure}
	\caption{Grid independence studies with (a) the BF and (b) the IB-CC methods considering surface heat fluxes for the inert isothermal 2-D cylinder case by Knight et al.~\cite{knight2012assessment}.}
	\label{fig:knight_inert_iso_grid}
\end{figure}

In~\autoref{fig:knight_inert_iso_heat_pres}, surface pressure and heat flux distributions are compared with the experimental measurements from Knight et al.~\cite{knight2012assessment} and also with the numerical simulations of Nompelis from the same publication.
All methods accurately predict the pressure distribution.
Heat flux predictions of the BF and the IB-CC methods are in very good agreement. They match the experimental measurements better than the numerical simulations of Nompelis. Slight differences in heat fluxes are expected to be due to the differences in grid resolutions at the surface.
Grid convergence studies have been carried out with both the BF and the IB-CC methods as summarized in~\autoref{tab:grid_conv} and showcased for the variation in heat fluxes in~\autoref{fig:knight_inert_iso_grid}. 
Four levels of resolution are considered with the minimum cell size at the surface approximately halving with each step. 
For both solvers, results obtained on the medium\nobreakdash-fine resolution mesh are considered grid converged, as they are essentially identical to the results obtained on the fine meshes.
For these grids, smallest cell size near the wall is $ 1.0 \times 10^{-7} $ m for the BF method and $ 6.25 \times 10^{-7} $ m for the IB-CC method.
An interesting observation is that IB-CC method under\nobreakdash-predicts the heat flux on coarse meshes, as intuitively expected, whereas the BF method over\nobreakdash-predicts the heat flux on coarse meshes. This difference in the convergence trend is a sign of complex interactions between transport and chemistry.

\begin{table}[h] 
	\centering
	\caption{Grid resolution and stagnation point details for the inert isothermal 2-D cylinder case, where $\Delta h_w$ is the effective wall-normal cell size at the wall, $p_{0}$ is the stagnation point pressure, and $q_{w,0}$ is the stagnation point wall heat flux.}
	\label{tab:grid_conv}
	\begin{tabular}{@{}llll@{}}
		\toprule
		Solver \& Grid Resolution & $\Delta h_w$ [\textmu m] & $p_{0}$ [kPa] & $q_{w,0}$ [MW/m$^2$]   \\ \midrule
            Experiment~\cite{knight2012assessment} & N/A & 52.26 $\pm$ 3.034 & 7.402 $\pm$ 0.220  \\ \midrule
            Nompelis~\cite{knight2012assessment} & 7.0\footnotemark & 52.40 & 5.971 \\ \midrule
		BF (coarse) & 0.44 & 54.38 & 8.422 \\
		BF (medium-coarse) & 0.22 & 53.29 & 7.576 \\ 
		BF (medium-fine) & 0.1 & 53.05 & 7.345 \\
		BF (fine) & 0.05 & 52.95 & 7.308 \\ \midrule
		IB-CC (coarse) & 2.5 & 51.57 & 6.684 \\
		IB-CC (medium-coarse) & 1.25 & 51.57 & 7.028 \\
		IB-CC (medium-fine) & 0.625 & 51.58 & 7.144 \\ 
		IB-CC (fine) & 0.3125 & 51.58 & 7.189 \\ \midrule 
            IB-GP~\cite{baskaya2022verification} & 1.0 & 52.83 & 0.167 \\
            \bottomrule
           \multicolumn{4}{l}{\footnotemark[1]\footnotesize{Personal communication with Ioannis Nompelis.} } 
	\end{tabular}
\end{table}

For this case, the IB-GP method is not able to predict the heat flux correctly. 
A similar underprediction has also been reported in literature for another ghost-cell IB\nobreakdash-AMR solver by Brahmachary et al.~\cite{brahmachary2021role}, where the issue has been linked to the reconstruction of temperature by linear interpolation. 
However, the cut\nobreakdash-cell IB method also resorts to second-order reconstruction schemes and can predict the heat flux correctly. Therefore, we attribute the observed deficiencies to conservation errors incurred through the ghost\nobreakdash-cell IB method.  
This hypothesis is further discussed in~\autoref{sec:GP_vs_CC}, where results obtained with two independently developed ghost\nobreakdash-cell IB methods are presented.

\begin{figure} [!h]
	\begin{subfigure}[t]{.49\textwidth}\centering
		\includegraphics[clip, trim=0cm 0cm 0cm 0cm, 
		width=0.999\columnwidth]{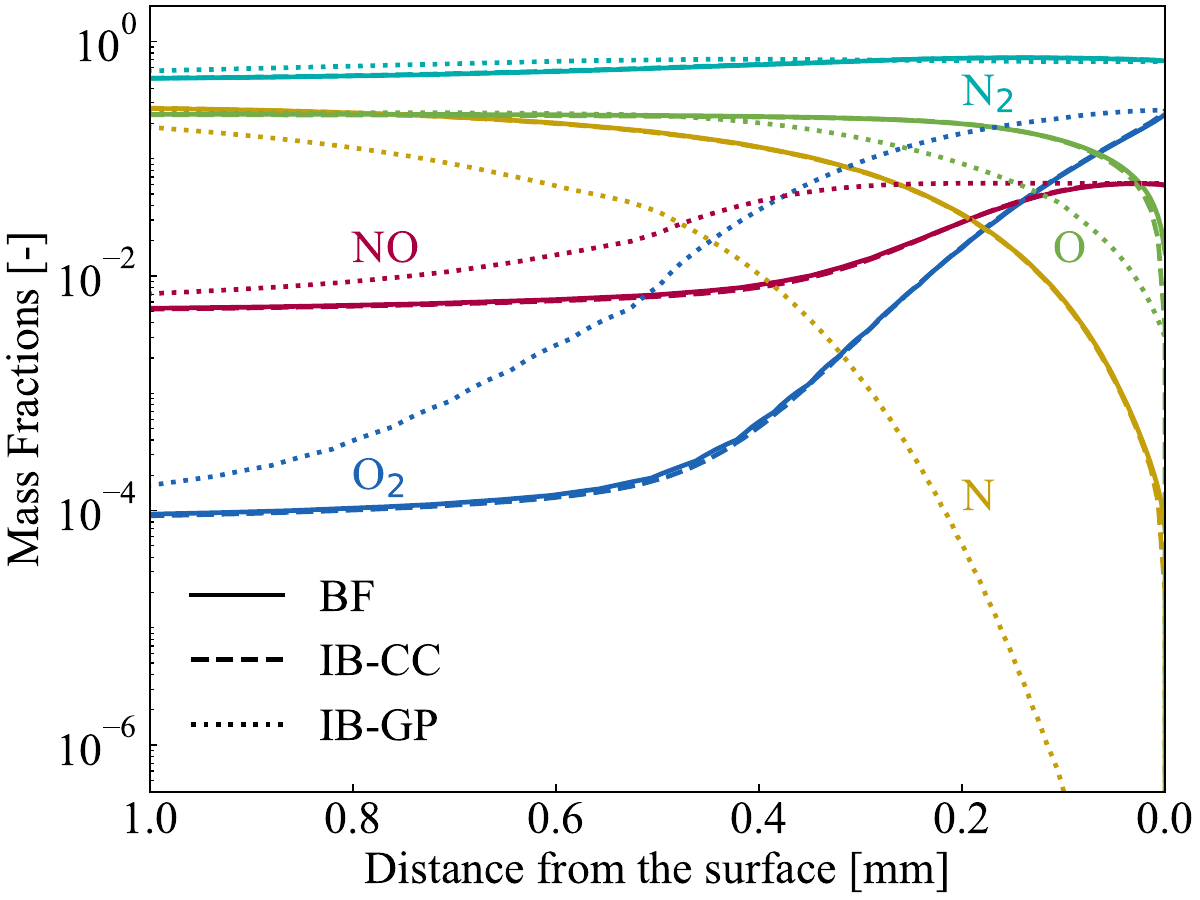}
		\caption{}
	\end{subfigure}
 \hspace{0.1cm}
	\begin{subfigure}[t]{.49\textwidth}\centering
		\includegraphics[clip, trim=0cm 0cm 0cm 0cm, 
		width=0.999\columnwidth]{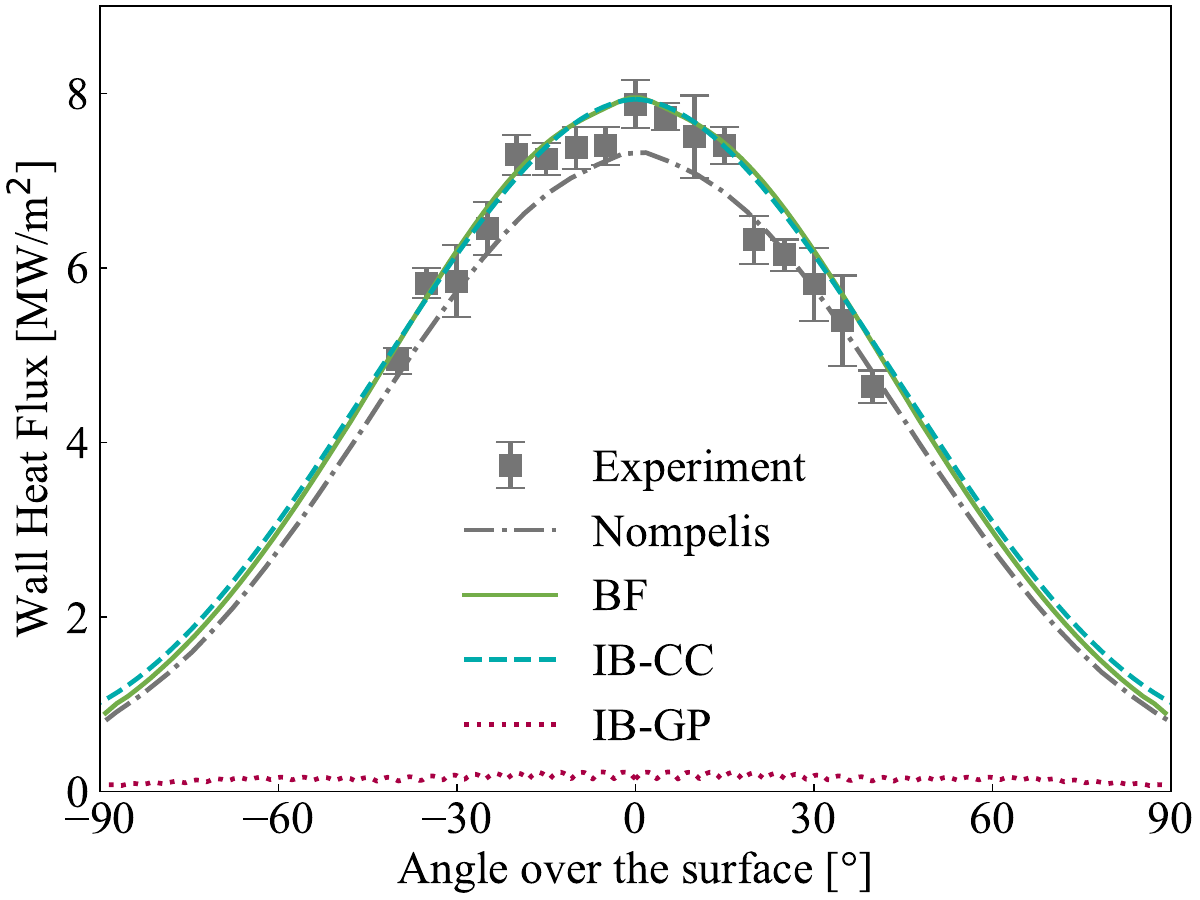}
		\caption{}
	\end{subfigure}
	\caption{Comparison of (a) mass fractions along the stagnation line and (b) heat fluxes over the surface for the catalytic isothermal 2-D cylinder case by Knight et al.~\cite{knight2012assessment}.}
	\label{fig:knight_cata_mass_heat}
\end{figure}

\subsubsection{Catalytic Isothermal Wall} \label{sec:knight_cata}

Exothermic catalytic reactions enhance the surface heat flux through a diffusive heat flux contribution.
A fully catalytic wall ($\gamma = 1.0$) at the same temperature of \SI{300}{\kelvin} is considered, as Karl et al.~\cite{karl2003high} state that this boundary condition is closest to what they have assumed for the experiments. It is, however, reasonable to assume that fully catalytic conditions were only achieved for a short duration at the beginning of the experiment. The fully catalytic boundary condition imposes the recombination of all atoms impinging on the surface, while still respecting the physical limits set by species diffusion.

The species mass fractions along the stagnation line and the total surface heat flux distributions are shown in~\autoref{fig:knight_cata_mass_heat}. 
Predictions of the BF and the IB-CC methods are in excellent agreement, both in terms of species concentrations and surface heat fluxes. Because the cold wall itself already promotes recombination reactions in the boundary layer, accounting for catalysis leads only to a minor increase in the heat fluxes, which remain within the experimental uncertainties. It is therefore difficult to draw conclusions on the effective value of the recombination coefficient in the experiment.

Another interesting observation could be made by comparing the level of agreement between the BF and the IB-CC results for the heat fluxes at the inert wall shown in~\autoref{fig:knight_inert_iso_heat_pres} and with the fully catalytic wall shown in~\autoref{fig:knight_cata_mass_heat}. 
Taking the BF method as reference, it is seen that at the stagnation point, results of the IB-CC method with the inert wall are 2.7\% lower, while with the fully catalytic wall the difference is only 0.2\%. It can be argued that this better agreement is mostly associated with the dominant nature of the catalytic boundary condition.
Our previous comments regarding the differences seen in the IB-GP method for the inert isothermal case apply here as well.

\begin{figure} [tb] \centering
	\begin{subfigure}[t]{.39\textwidth}\centering
		\includegraphics[clip, trim=0.75cm 0.75cm 13.5cm 3cm, 
		width=0.899\columnwidth]{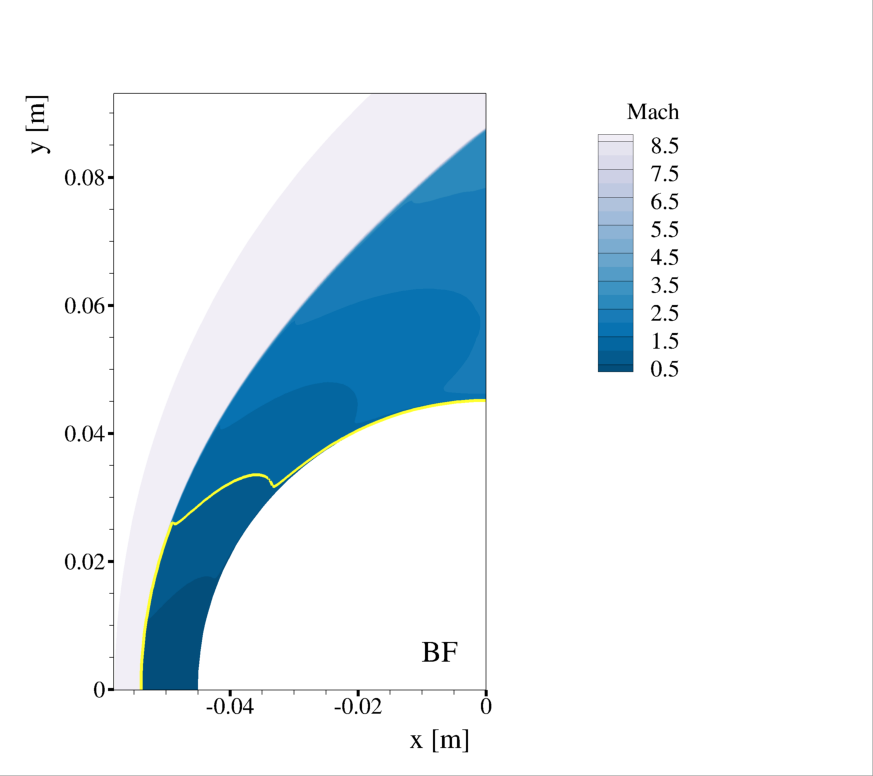}
        \caption{}
	\end{subfigure}
	\begin{subfigure}[t]{.39\textwidth}\centering
		\includegraphics[clip, trim=0.75cm 0.75cm 13.5cm 3cm, 
		width=0.899\columnwidth]{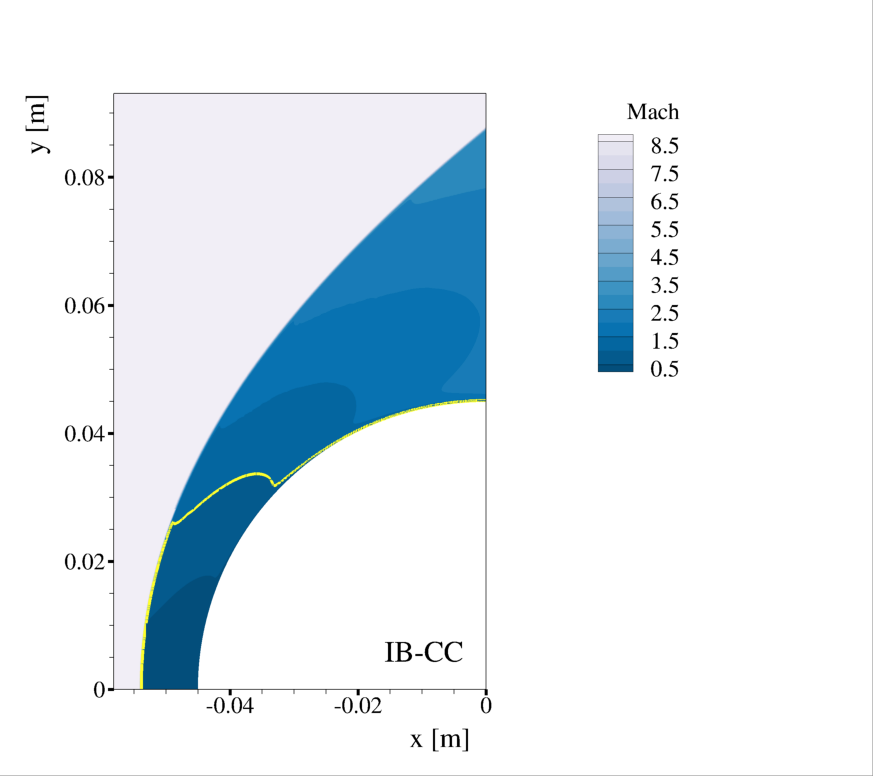}
        \caption{}
	\end{subfigure}
    \begin{subfigure}[t]{.19\textwidth}\centering
		\includegraphics[clip, trim=19cm 5cm 5cm 3cm, 
		width=0.75\columnwidth]{contour_plots/INCA_cata_M_new.png}
	\end{subfigure}
	\caption{Comparison of Mach number contours for the catalytic isothermal 2-D cylinder case by Knight et al.~\cite{knight2012assessment} obtained with (a) the BF and (b) the IB-CC methods. The sonic line is indicated with the bright yellow line.}
	\label{fig:contour_Knight_M}
\end{figure}

\begin{figure} [h] \centering
	\begin{subfigure}[t]{.39\textwidth}\centering
		\includegraphics[clip, trim=0.75cm 0.75cm 13.5cm 3cm, 
		width=0.899\columnwidth]{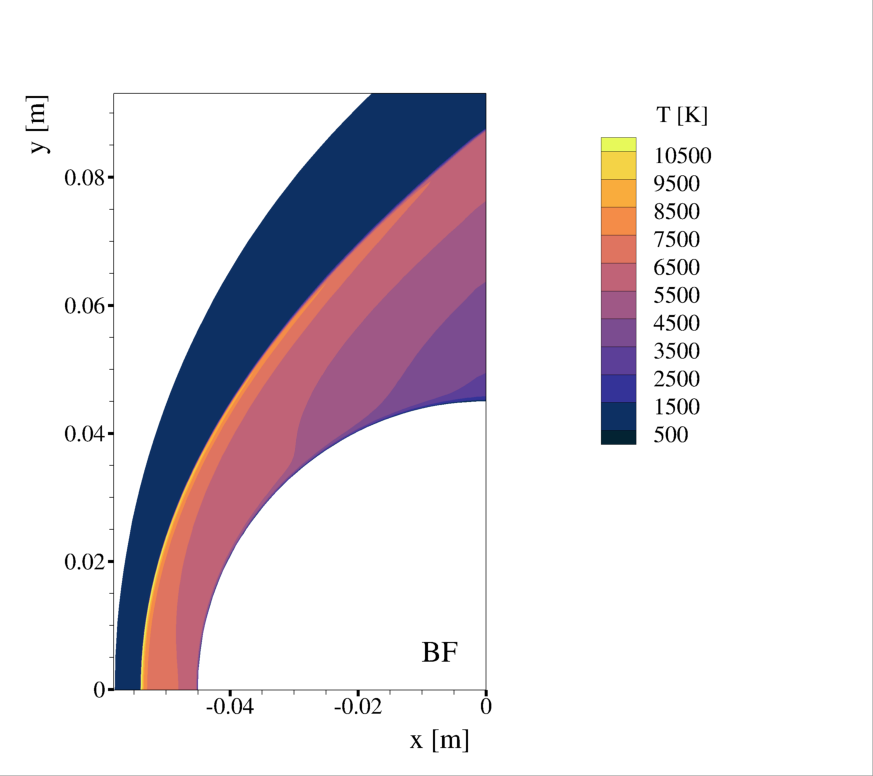}
        \caption{}
	\end{subfigure}
	\begin{subfigure}[t]{.39\textwidth}\centering
		\includegraphics[clip, trim=0.75cm 0.75cm 13.5cm 3cm, 
		width=0.899\columnwidth]{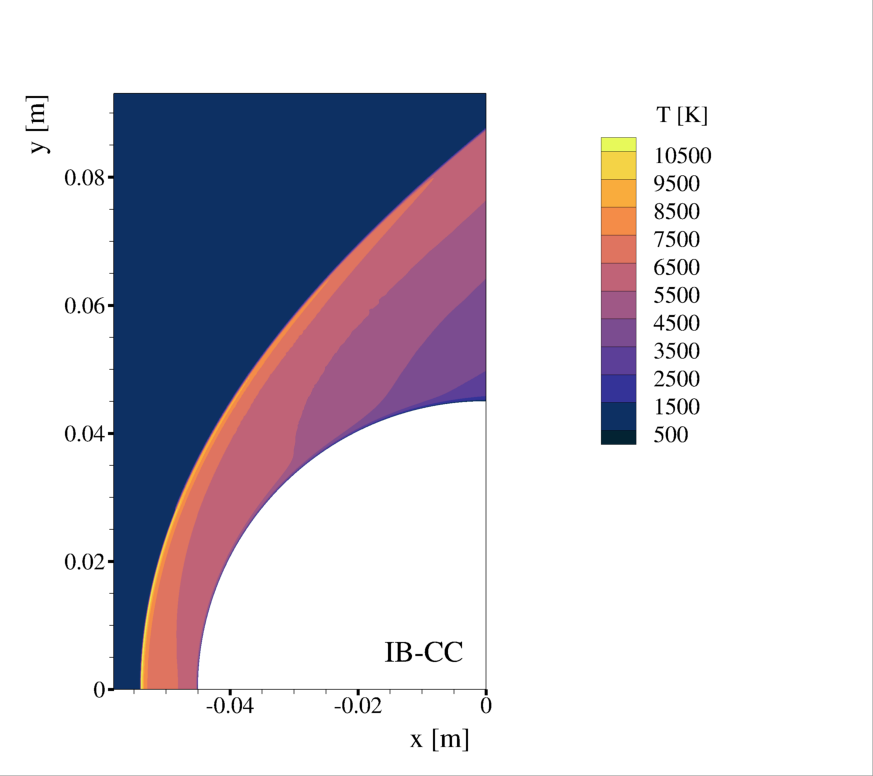}
        \caption{}
	\end{subfigure}
    \begin{subfigure}[t]{.19\textwidth}\centering
		\includegraphics[clip, trim=19cm 5cm 5cm 3cm, 
		width=0.75\columnwidth]{contour_plots/INCA_cata_T_new.png}
	\end{subfigure}
	\caption{Comparison of temperature contours for the catalytic isothermal 2-D cylinder case by Knight et al.~\cite{knight2012assessment} obtained with (a) the BF and (b) the IB-CC methods.}
	\label{fig:contour_Knight_T}
\end{figure}

\begin{figure} [!h] \centering
	\begin{subfigure}[t]{.39\textwidth}\centering
		\includegraphics[clip, trim=0.75cm 0.75cm 13.5cm 3cm, 
		width=0.899\columnwidth]{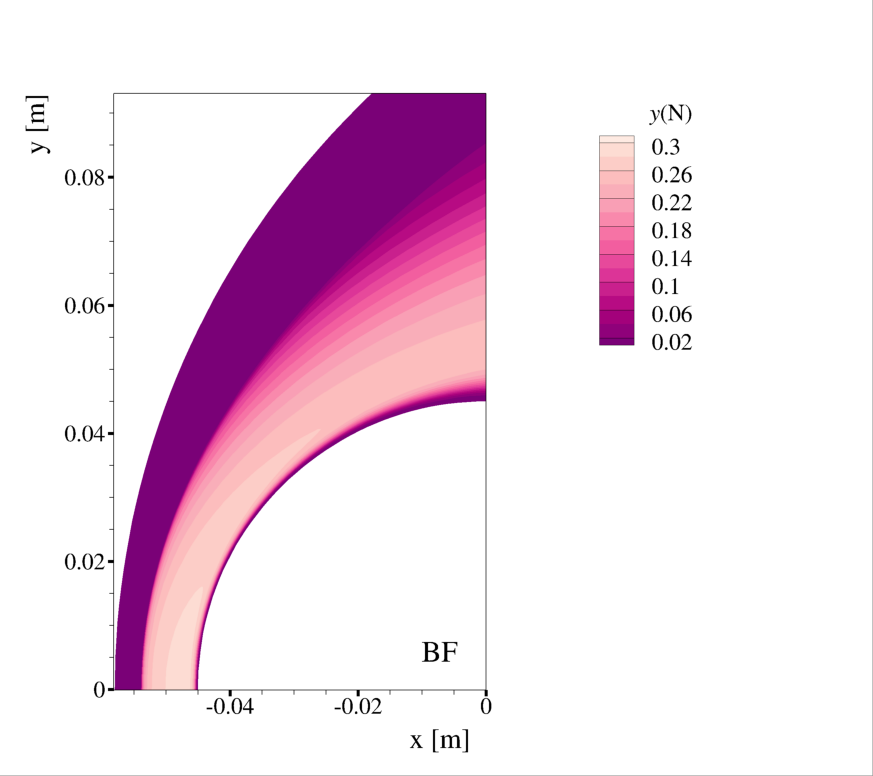}
        \caption{}
	\end{subfigure}
	\begin{subfigure}[t]{.39\textwidth}\centering
		\includegraphics[clip, trim=0.75cm 0.75cm 13.5cm 3cm, 
		width=0.899\columnwidth]{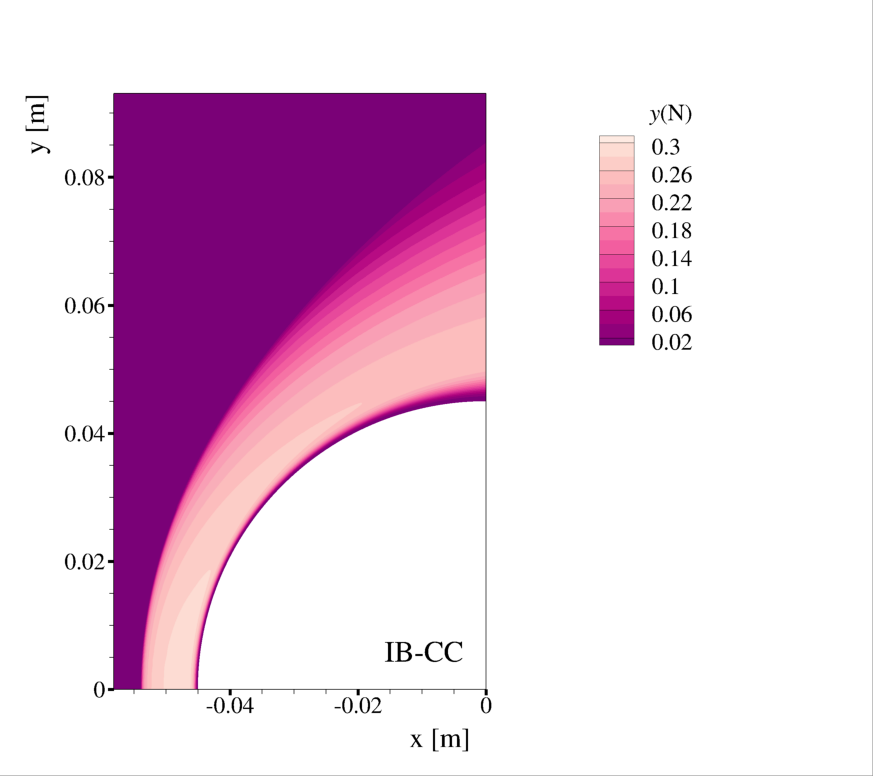}
        \caption{}
	\end{subfigure}
    \begin{subfigure}[t]{.19\textwidth}\centering
		\includegraphics[clip, trim=19cm 5cm 5cm 3cm, 
		width=0.75\columnwidth]{contour_plots/INCA_cata_N_new.png}
	\end{subfigure}
	\caption{Comparison of atomic nitrogen mass fraction contours for the catalytic isothermal 2-D cylinder case by Knight et al.~\cite{knight2012assessment} obtained with (a) the BF and (b) the IB-CC methods.}
	\label{fig:contour_Knight_N}
\end{figure}

To complete the analysis, contour plots are presented for Mach numbers in~\autoref{fig:contour_Knight_M}, for temperatures in~\autoref{fig:contour_Knight_T}, and for atomic nitrogen concentrations in~\autoref{fig:contour_Knight_N}. These contour plots further confirm the preceding quantitative discussions by once again reflecting the excellent agreement between the BF and the IB-CC methods. From the trace of the sonic line, to the peak shock temperatures, and to the extent of the species boundary layer marked by nitrogen accumulation, the results are in perfect agreement.


\subsection{2-D Ablator} \label{sec:abla}
To validate the ablative boundary condition implementation and to assess the IB methods for GSI, a subsonic plasma wind tunnel experiment conducted at the von Karman Institute for Fluid Dynamics (VKI) by Helber et al.~\cite{helber2017determination} is considered. The experiment exposes a graphite sample with a hemispherical nose of radius \SI{25}{\milli\meter} and a downstream extension of \SI{250}{\milli\meter} to nitrogen plasma. The sample undergoes ablation through nitridation reactions
\begin{equation}
    	\text{C}_{(\text{solid})} + \text{N} \rightarrow \text{CN} \:,
\end{equation}
according to Eqs.~(\ref{eq:mass_balance}-\ref{eq:vblow}) with the nitridation efficiency coefficient 
\begin{equation} \label{eq:gamma}
    \gamma = 7.91 \cdot 10^{-2} \exp \left( -\frac{5663}{T_\text{wall}} \right) \:.
\end{equation}
The nitridation efficiency was calibrated based on these particular experiments~\cite{helber2017determination}.
The simulations discussed in the following include mass blowing due to ablation, but do not account for the very slow shape change of the sample.

First, we reproduce the experiment numerically using the BF method.
For these simulations, a 9-species nitrogen\nobreakdash-carbon mixture is considered, including free electrons and ionized species.
These simulations yielded a stagnation point mass blowing rate of $3.41$ \SI{}{\gram/\square\meter\second}, which is within the experimental uncertainty range set by $2.8864 \mp 0.965$ \SI{}{\gram/\square\meter\second}. This validates the ablation model based on~\autoref{eq:gamma}.

Having confidence in the ablation model and its implementation in the BF method, the experimental test case is simplified to a 2-D geometry without ionized species to reduce the computational cost and to avoid straying too far from the objective of evaluating immersed boundary methods for an ablative boundary condition. 
Freestream conditions of this 2-D case are given in~\autoref{tab:abla}. A 6-species mixture of $ [ \text{N}_2, \text{N}, \text{CN}, \text{C}_3, \text{C}_2, \text{C} ] $ is considered with chemical mechanisms from Olynick et al.~\cite{olynick1999aerothermodynamics}. For all methods, the grid resolution at the wall is $1 \times 10^{-5} $ m in the wall\nobreakdash-normal direction.

\begin{table}[h]
	\centering
	\caption{Freestream conditions for the 2-D ablator case.}
	\label{tab:abla}
	\begin{tabular}{@{}cccccc@{}}
		\toprule
		$ u_\infty $ [m/s] & $ T_\infty $ [K] & $ T_\text{wall} $ [K] & $ p_\infty $ [Pa]  &  $ y(\text{N}_2) $  & $ y(\text{N}) $  \\ \midrule
		1570 & 10280 & 2407 & 1500 & 9.77659e-05 & 0.9999022341 \\ \bottomrule
	\end{tabular}
\end{table}

\begin{figure} [tb]
	\begin{subfigure}[t]{.49\textwidth}\centering
		\includegraphics[clip, trim=0cm 0cm 0cm 0cm, 
		width=0.999\columnwidth]{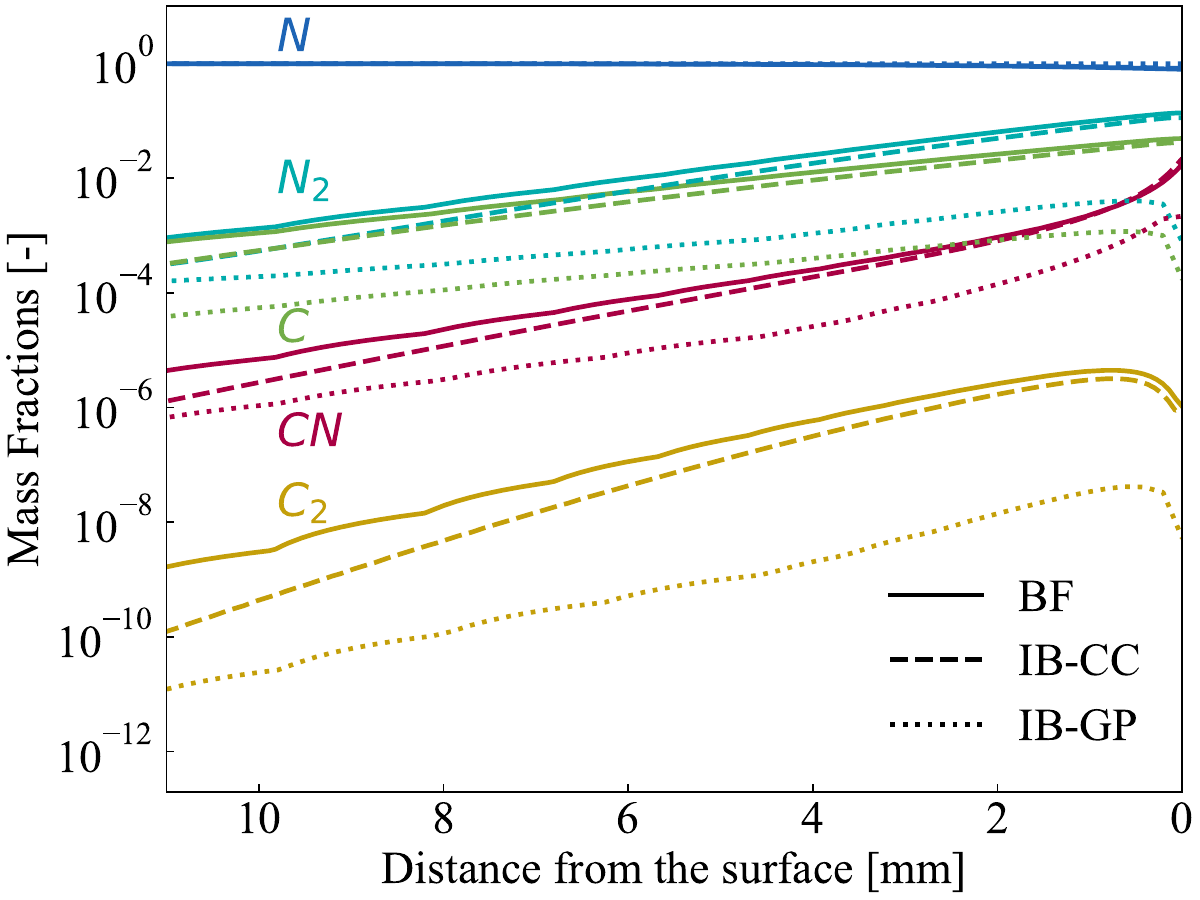}
		\caption{}
	\end{subfigure}
 \hspace{0.1cm}
	\begin{subfigure}[t]{.49\textwidth}\centering
		\includegraphics[clip, trim=0cm 0cm 0cm 0cm, 
		width=0.999\columnwidth]{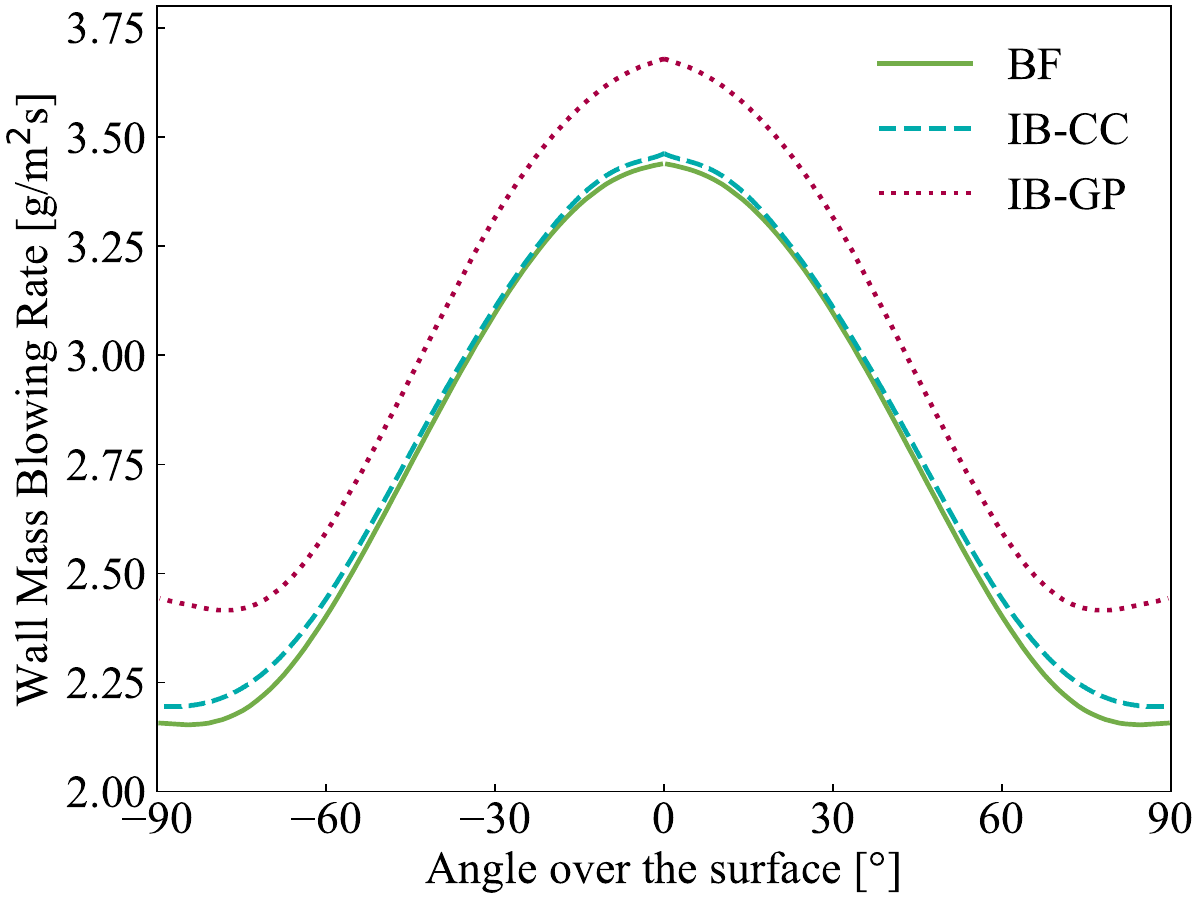}
		\caption{}
	\end{subfigure}
	\caption{Comparison of (a) mass fractions along the stagnation line and (b) mass blowing rates over the surface for the 2-D ablator case.}
	\label{fig:ablator}
\end{figure}

The mass fractions along the stagnation line and the mass blowing rates over the wall are shown in \autoref{fig:ablator}. Mass fractions for $\text{C}_3$ are not seen as they are almost zero.
Predictions of the BF and the IB-CC methods agree well with each other. 
Overall, the production of CN at the wall and the dissociation of it through gas-phase reactions to form atomic nitrogen are well captured.
Mass blowing rates from the BF and the IB-CC methods are also in very good agreement.
The IB-GP method show noticeable discrepancies for the mass fractions along the stagnation line and for the surface mass blowing rates. 
Despite the apparent quantitative mismatch, also the IB-GP method captures the profiles qualitatively well in the absence of strong gradients near the wall. 

\begin{figure} [h] \centering
    \begin{minipage}{0.75\linewidth} \centering
        \begin{subfigure}[t]{.9\textwidth}\centering
		\includegraphics[clip, trim=0.75cm 0.75cm 3cm 17.5cm, 
		width=0.899\columnwidth]{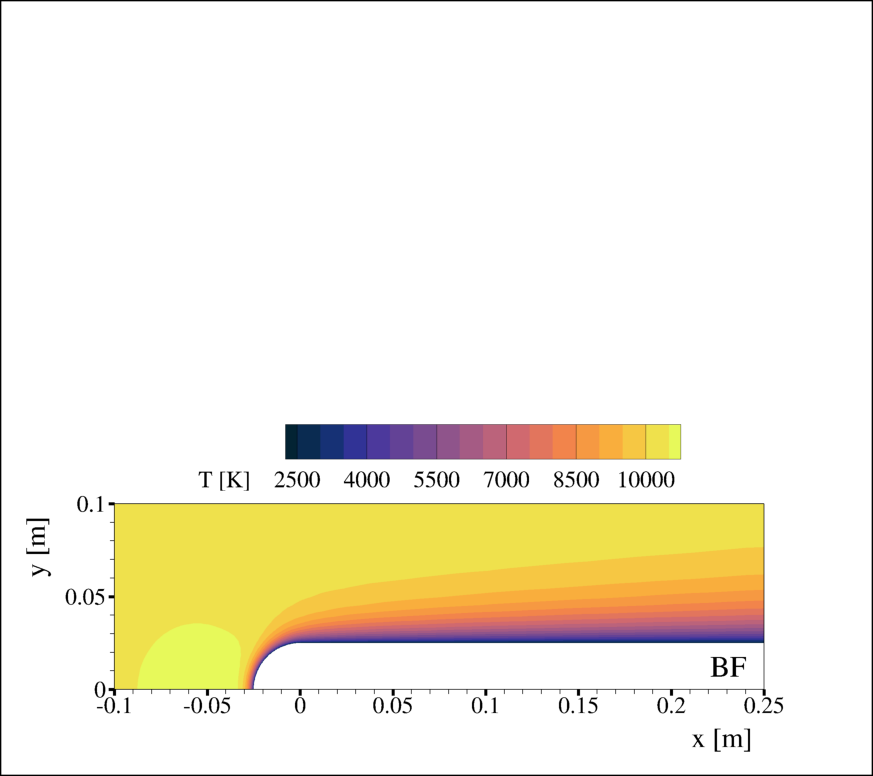}
    \put(-265,70){(a)}
	\end{subfigure}
	\begin{subfigure}[t]{.9\textwidth}\centering
		\includegraphics[clip, trim=0.75cm 0.75cm 3cm 17.5cm, 
		width=0.899\columnwidth]{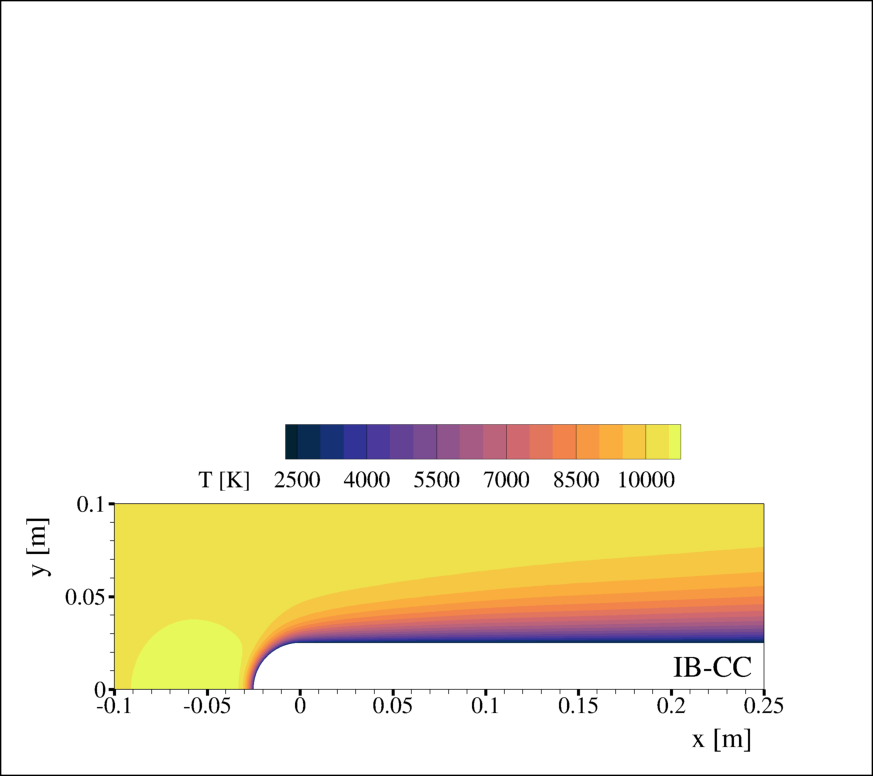}
    \put(-265,70){(b)}
	\end{subfigure}
    \end{minipage}
    \begin{minipage}{0.19\linewidth} \centering
    \begin{subfigure}[t]{.99\textwidth}\centering
		\includegraphics[clip, trim=13cm 10cm 11cm 0.2cm, 
		width=0.72\columnwidth]{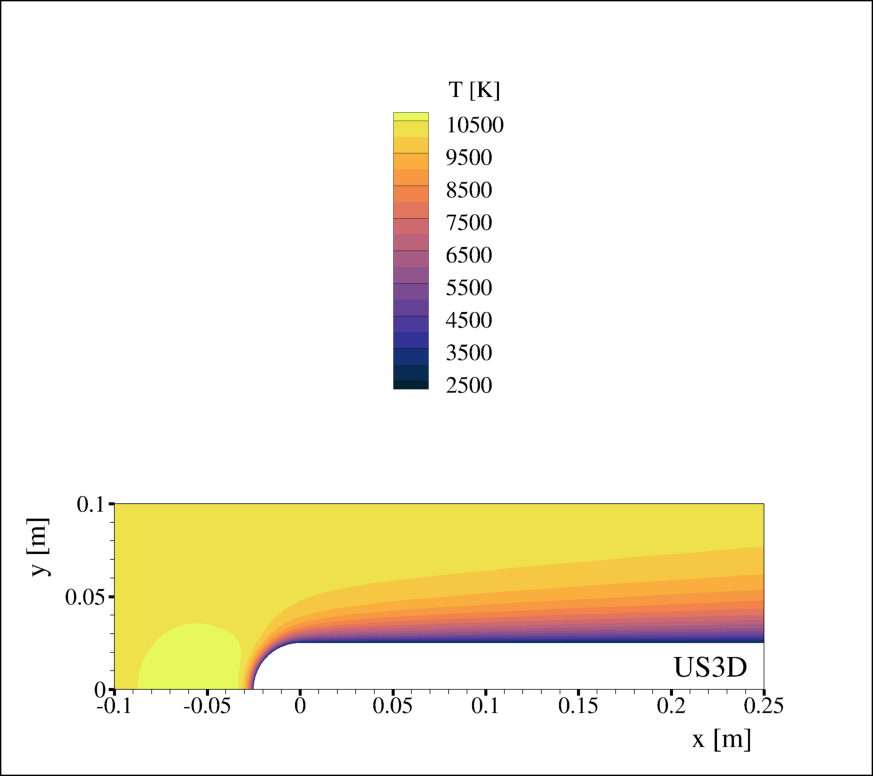}
	\end{subfigure}
    \end{minipage}
	\caption{Comparison of the temperature contours for the 2-D ablator case obtained with (a) the BF and (b) the IB-CC methods.}
	\label{fig:contour_abla_T}
\end{figure}

\begin{figure} [!h] \centering
    \begin{minipage}{0.75\linewidth} \centering
        \begin{subfigure}[t]{.9\textwidth}\centering
		\includegraphics[clip, trim=0.75cm 0.75cm 3cm 17.5cm, 
		width=0.899\columnwidth]{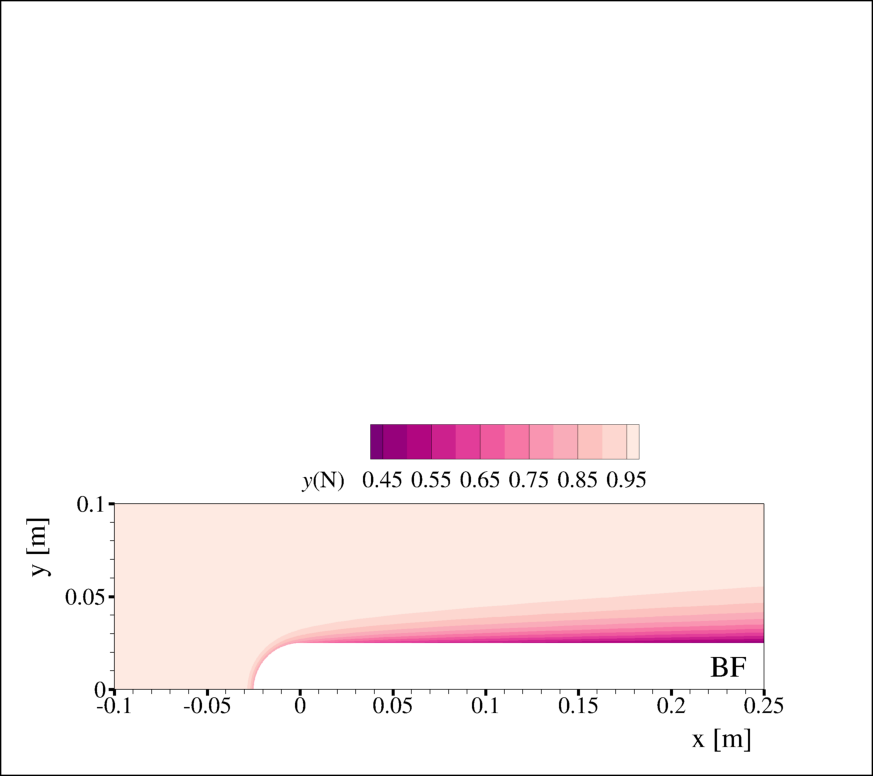}
  \put(-265,70){(a)}
	\end{subfigure}
	\begin{subfigure}[t]{.9\textwidth}\centering
		\includegraphics[clip, trim=0.75cm 0.75cm 3cm 17.5cm, 
		width=0.899\columnwidth]{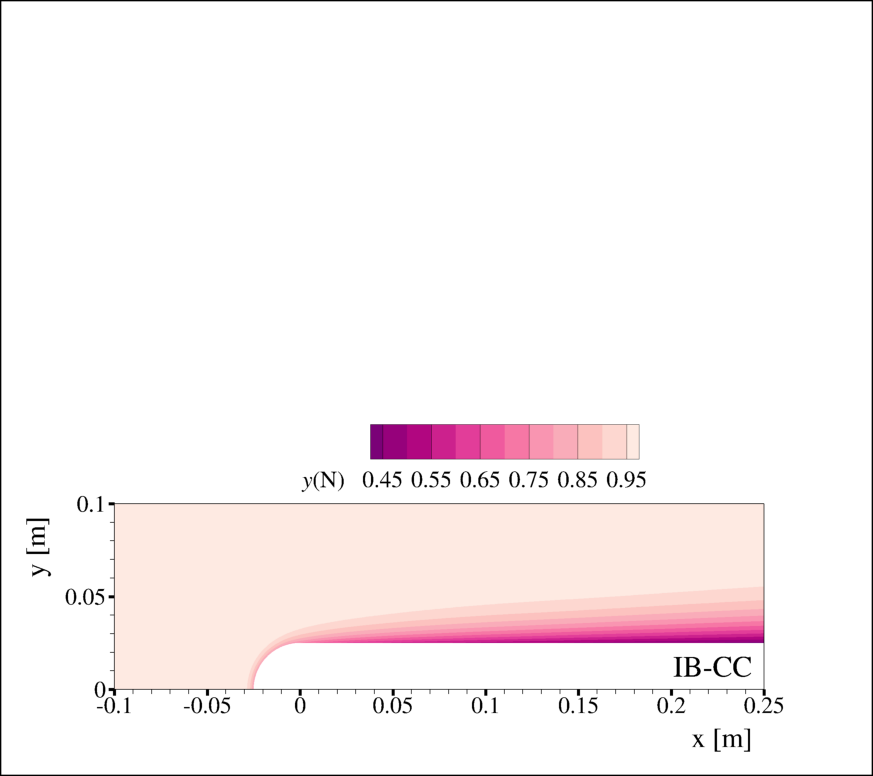}
  \put(-265,70){(b)}
	\end{subfigure}
    \end{minipage}
    \begin{minipage}{0.19\linewidth} \centering
    \begin{subfigure}[t]{.99\textwidth}\centering
		\includegraphics[clip, trim=13cm 10cm 11cm 0.2cm, 
		width=0.72\columnwidth]{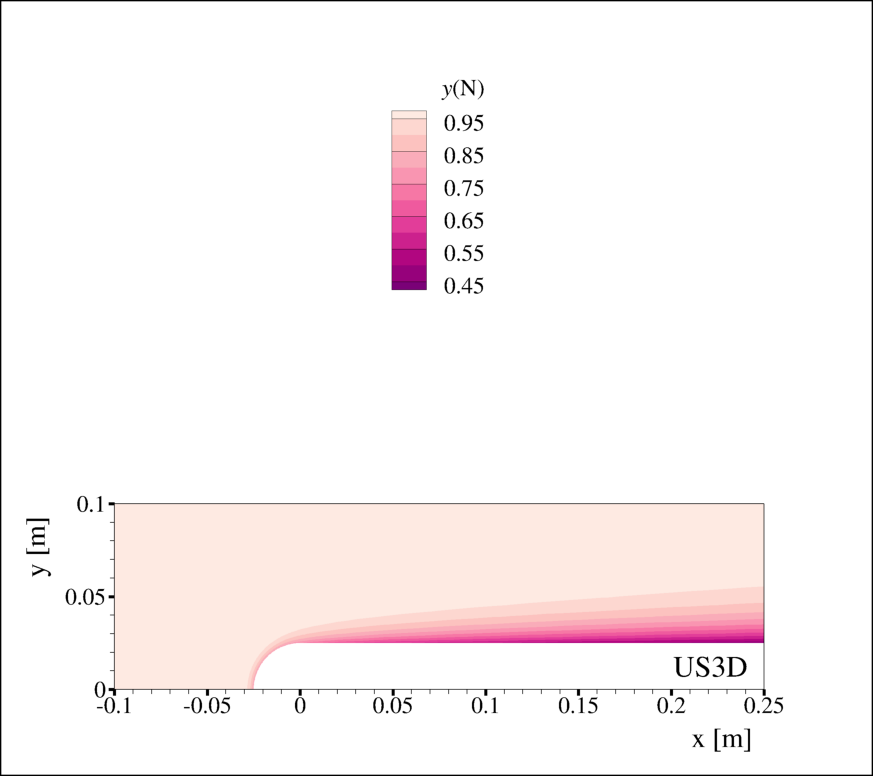}
	\end{subfigure}
    \end{minipage}
	\caption{Comparison of the atomic nitrogen mass fraction contours for the 2-D ablator case obtained with (a) the BF and (b) the IB-CC methods.}
	\label{fig:contour_abla_N}
\end{figure}

Temperature and atomic nitrogen contours for the simulations with the BF and the IB-CC methods are shown in Figs.~\ref{fig:contour_abla_T} and~\ref{fig:contour_abla_N}. Results of both methods agree very well on the thermal gradient over the surface and on the recombination of nitrogen as temperature drops.


\section{On the Importance of Conservative Boundary Conditions}
\label{sec:GP_vs_CC}

In the previous section, we have demonstrated that INCA with its cut\nobreakdash-cell IB method on block\nobreakdash-Cartesian AMR meshes performs on par with the reference solver US3D employing body\nobreakdash-fitted meshes. The ghost\nobreakdash-cell IB method of CHESS yields similar accuracy for cases with adiabatic walls; however, it cannot predict the heat flux at strongly cooled walls. We have attributed these inaccuracies to mass conservation errors as this is the most striking difference between ghost\nobreakdash-cell methods and the strictly conservative cut\nobreakdash-cell approach. However, the three solvers clearly differ also in several other aspects, such as the numerical schemes used for advection and diffusion driving forces. 
It is therefore unclear whether the observed deficiencies are inherent to the ghost-cell method or a particular implementation. 
To further corroborate the superiority of a conservative cut-cell (or cut-element) IB methodology, we have also applied the ghost\nobreakdash-cell method of INCA for selected cases.
By switching off the special flux treatment in cut-cells~\cite{baskaya3AF} employed in the preceding sections, a standard ghost\nobreakdash-cell method is obtained that only relies on the extrapolated fluid solutions near the boundary as described in~\autoref{sec:num_meth}.
Therefore, mass, momentum, and energy conservation are not exactly satisfied. This ghost\nobreakdash-cell method has nominally the same order of convergence as the conservative cut\nobreakdash-element method of INCA. 

\begin{figure} [tb]
	\begin{subfigure}[t]{.49\textwidth}\centering
		\includegraphics[clip, trim=0cm 0cm 0cm 0cm, 
		width=0.999\columnwidth]{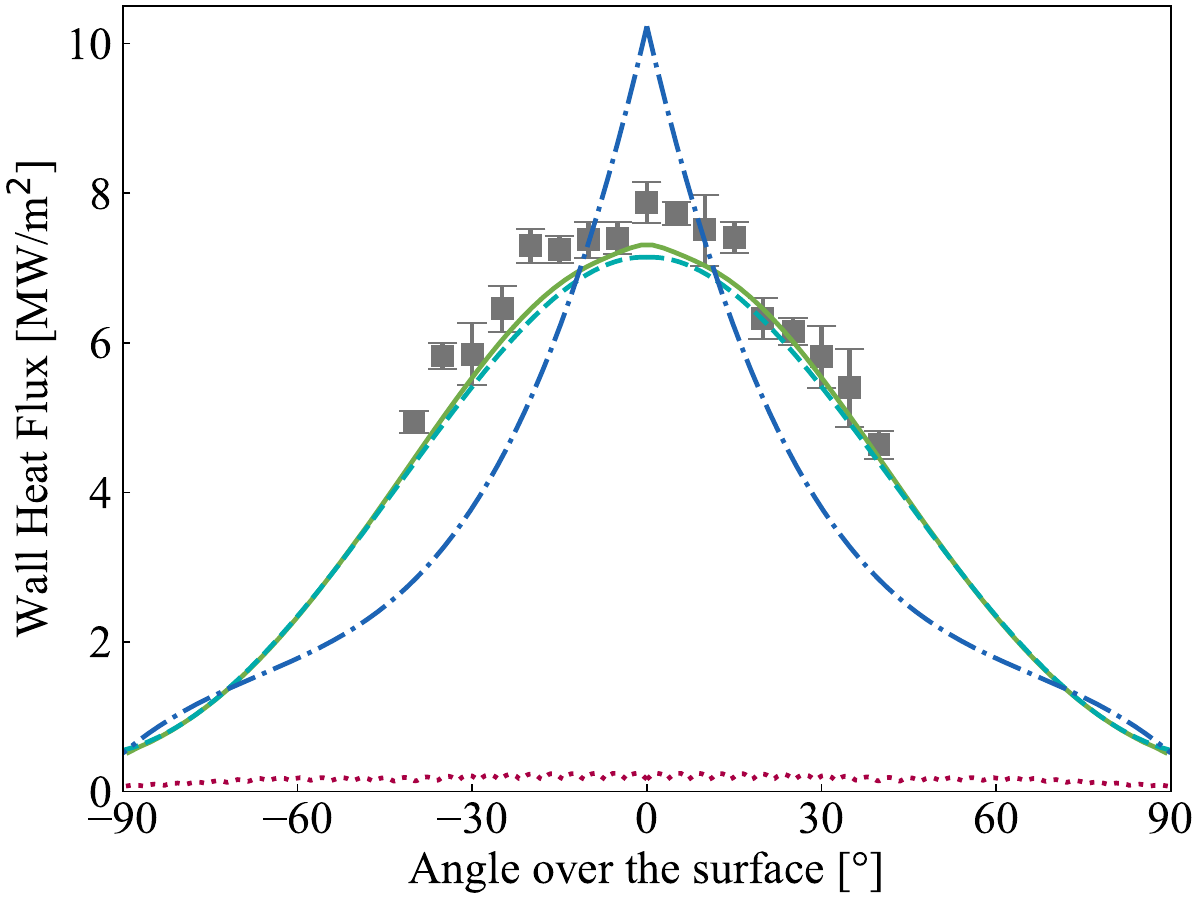}
		\caption{}
	\end{subfigure}
 \hspace{0.1cm}
	\begin{subfigure}[t]{.49\textwidth}\centering
		\includegraphics[clip, trim=0cm 0cm 0cm 0cm, 
		width=0.999\columnwidth]{from_pylotron/2D_ablator_fig_mdot_GP_py.pdf}
		\caption{}
	\end{subfigure}
	\caption{Comparison of (a) heat fluxes over the surface for the isothermal 2-D cylinder case by Knight et al.~\cite{knight2012assessment} and (b) mass blowing rates over the surface for the 2-D ablator case. ``INCA-CC'' refers to INCA with the cut\nobreakdash-cell based IB method, ``INCA-GP'' refers to the ghost\nobreakdash-cell IB method of INCA, and ``CHESS-GP'' refers to the ghost-cell IB method of CHESS~\cite{baskaya2022verification}.}
	\label{fig:CC_vs_GP}
\end{figure}

The comparison of the two methods is shown for the two most challenging benchmark cases in~\autoref{fig:CC_vs_GP}, where the previous cut\nobreakdash-cell based results from the INCA solver are denoted by ``INCA-CC'' and the ghost\nobreakdash-cell based results are denoted by ``INCA-GP''. 
Similarly, results with the ghost-cell IB method of the CHESS solver are denoted by ``CHESS-GP''.
It can be seen that regardless of the various 
differences between INCA and CHESS,
in both cases the ghost\nobreakdash-cell methods are unsuccessful in predicting the surface heat fluxes and the mass blowing rates.
For the 2-D cylinder case with an isothermal wall,
the heat flux prediction of the INCA-GP method is closer in magnitude to the INCA-CC method and to the body-fitted reference from US3D than to the results obtained with the CHESS ghost-cell IB method; however, both ghost-cell methods give clearly wrong results.
For the ablative case, the ghost\nobreakdash-cell based method of INCA yields a very similar overprediction as the IB method of CHESS.

Comparable inaccuracies are observed for two independently developed ghost-cell IB methods.
The only difference between the ghost-cell IB method of INCA and INCA's cut-cell method, which shows excellent agreement with the reference data, is the conservative flux treatment.
This further consolidates the diagnosis that conservation errors inherent to ghost\nobreakdash-cell IB methods are responsible for large errors at cold walls. It is expected that these conservation errors converge at the same rate as the truncation errors of the baseline schemes. That is, conservation errors are very small unless gradients of the conservative variables are very large. This explains why errors manifest at cold walls and not at adiabatic walls.

\section{Conclusion} \label{sec:conc}

We have evaluated the accuracy of immersed boundary methods for atmospheric entry conditions, including the influence of chemical nonequilibrium and gas-surface interactions. 
The benchmark cases have considered the accurate modelling of gas chemistry, mass diffusion, surface catalysis, and surface mass blowing due to ablation.

Computational results obtained with the cut-element IB method in the AMR solver INCA are in an almost perfect agreement with the reference data for all considered cases, and as accurate as the results obtained with US3D on body-fitted meshes. Particularly for surface heat flux and mass blowing rate predictions, the benefit of an IB method that strictly conserves mass, momentum, and energy, such as the cut\nobreakdash-element method in INCA, is clearly demonstrated in this study.
After comparing this method with two non-conservative ghost\nobreakdash-cell methods implemented in INCA and an independently developed solver, we saliently remark that numerical anomalies causing mispredictions of sensitive surface quantities can occur when using non-conservative IB formulations.

CFD solvers that provide automatic mesh generation and adaptation to represent detailed and moving geometries with IB methods have many promising advantages, but the accuracy of the numerical schemes used for predicting surface quantities must be analyzed rigorously before they can be used for predictive simulations. 
The selection of a set of well\nobreakdash-defined test cases by mutual collaboration between research groups is crucial in converging to a robust consensus on the prediction of these surface states.
To that end, this paper establishes such a set of fundamental benchmark cases with reacting surfaces, which can be used for the verification and validation of hypersonic flow solvers, while assessing the accuracy of immersed boundary methods for atmospheric entry applications.

\section*{Acknowledgments}
The authors would like to thank Dr.~Davide Ninni, Dr.~Francesco Bonelli, and Prof.~Giuseppe Pascazio from Politecnico di Bari for their collaboration and discussions on the results.
From TU Delft, the authors would like to thank Prof.~Georg Eitelberg for his insight regarding the experiments conducted at DLR and Dr.~Ferdinand Schrijer for his comments on the manuscript. 
We also thank the Delft High Performance Computing Centre for providing access to DelftBlue and SURF (www.surf.nl) for the support in using the National Supercomputer Snellius.

\appendix

\hypertarget{sec:appendix}{\section*{Appendix}}

\subsection*{Analytical Solution of the 1-D Catalytic Diffusion Problem} \label{sec:cata_derivation}

Following the derivation proposed by Bariselli~\cite{art:bariselli2016}, substituting Fick's law into the molar continuity equation, and solving for the zero-advection,
constant temperature, steady-state solution one obtains
\begin{equation}
    \nabla \cdot \left( n \frac{M_\text{N}}{\overline{M}} D_{\text{N}_2,\text{N}} \nabla (x_{\text{N}_2})\right) = 0 \:,
\end{equation}
with $n$ as the number density.
For the current binary mixture $M_{N_2} = 2 M_{N}$ and $ \overline{M} = \sum_i x_i M_i $, which in 1-D leads to
\begin{equation}
    \frac{d}{d\eta} \left( \frac{1}{x_{\text{N}_2} + 1} \left( \frac{d}{d\eta} x_{\text{N}_2}\right) \right) = 0 \:,
\end{equation}
with $ \eta $ as the spatial coordinate. Solving for $ x_{\text{N}_2} $ yields
\begin{equation}
    x_{\text{N}_2} = \frac{e^{C_1 M_\text{N} \eta}e^{C_2 M_\text{N}}}{M_\text{N}} \:,
\end{equation}
with $ C_1 $ and $ C_2 $ as integration constants to be found through the boundary conditions. Firstly, by knowing that $ \left(X_{\text{N}_2}\right)_{\eta =0} = 0 $ at the free-stream reservoir
\begin{equation}
    C_2 = \frac{\ln M_\text{N}}{M_\text{N}} \:.
\end{equation}
Secondly, by equating the diffusion flux to the chemical production rate at the wall, $ (J_{\text{N}_2} = \Dot{\omega}_{\text{N}_2})_{\eta=L} $, which gives
\begin{equation}
    \left(\frac{C_1 M_\text{N}}{2 - e^{C_1 L M_\text{N}}} = \frac{\gamma_\text{N}}{ 2 D_{\text{N}_2\text{N}}}\sqrt{\frac{k_B T}{2 \pi m_\text{N}}}\right)_{\eta = L} \:,
\end{equation}
where $k_B$ is the Boltzmann constant. The last expression can be solved iteratively through the Newton-Raphson method. The solution describes the species distribution as a function of spatial variable $\eta$.

 \bibliographystyle{elsarticle-num} 
 \bibliography{elsarticle-template-num}

\end{document}